\documentclass[aps,prl,floatfix,onecolumn,reprint,amsmath,amssymb,superscriptaddress,longbibliography]{revtex4-2}
\usepackage{lipsum}
\usepackage{amsfonts}
\usepackage{mathrsfs}
\usepackage{amsmath}% needed for subequations
\usepackage{color}
\usepackage{natbib}
\usepackage{graphicx}
\usepackage{bm}% bold maths
\usepackage{amssymb}
\usepackage{stmaryrd}
\usepackage{xspace}
\usepackage{epstopdf}
\usepackage{dcolumn}% Align table columns on decimal point
\usepackage{longtable}
\usepackage{multirow}
\usepackage{bbding}
\usepackage[colorlinks=true, letterpaper=true, pdfstartview=FitV, linkcolor=blue, citecolor=blue, urlcolor=blue]{hyperref}

\makeatletter

\newcommand{\Rmnum}[1]{\expandafter\@slowromancap\romannumeral #1@}
\makeatother
\begin{document}

%\title{Engineering Multiple Gapless Quantum Spin Hall States from Altermagnetic Multilayers}
%\title{Engineering Quantum Spin Hall Effects with Tunable Gapless Edge States \\ in Multilayer Altermagnets}

%\title{Intact Bulk-Boundary Correspondence in Altermagnetic Quantum Spin Hall Insulator Multilayers} 
\title{Quantum Spin Hall Effect with Extended Topologically Protected Features in Altermangetic Multilayers} 

%\title{Resumptive Bulk-Boundary Correspondence in Altermagnetic Quantum Spin Hall Insulator Multilayers}

\author{Zhiyu Chen}
\affiliation{Institute for Structure and Function $\&$ Department of Physics $\&$ Chongqing Key Laboratory for Strongly Coupled Physics, Chongqing University, Chongqing 400044, P. R. China}

\author{Fangyang Zhan}
\email{zhan\_fyang@cqu.edu.cn}
\affiliation{Institute for Structure and Function $\&$ Department of Physics $\&$ Chongqing Key Laboratory for Strongly Coupled Physics, Chongqing University, Chongqing 400044, P. R. China}
\affiliation{College of Mathematics and Statistics, Chongqing University, Chongqing 400044, P. R. China}

\author{Zheng Qin}
\affiliation{Institute for Structure and Function $\&$ Department of Physics $\&$ Chongqing Key Laboratory for Strongly Coupled Physics, Chongqing University, Chongqing 400044, P. R. China}

\author{Da-Shuai Ma}
\affiliation{Institute for Structure and Function $\&$ Department of Physics $\&$ Chongqing Key Laboratory for Strongly Coupled Physics, Chongqing University, Chongqing 400044, P. R. China}
\affiliation{Center of Quantum materials and devices, Chongqing University, Chongqing 400044, P. R. China}

\author{Dong-Hui Xu}
\email{donghuixu@cqu.edu.cn}
\affiliation{Institute for Structure and Function $\&$ Department of Physics $\&$ Chongqing Key Laboratory for Strongly Coupled Physics, Chongqing University, Chongqing 400044, P. R. China}
\affiliation{Center of Quantum materials and devices, Chongqing University, Chongqing 400044, P. R. China}

\author{Rui Wang}
\email{rcwang@cqu.edu.cn}
\affiliation{Institute for Structure and Function $\&$ Department of Physics $\&$ Chongqing Key Laboratory for Strongly Coupled Physics, Chongqing University, Chongqing 400044, P. R. China}
\affiliation{Center of Quantum materials and devices, Chongqing University, Chongqing 400044, P. R. China}
\date{\today}

\begin{abstract}

Conventional topological classification theory dictates that time-reversal symmetry confines the quantum spin Hall (QSH) effect to a $\mathbb{Z}_2$ classification, permitting only a single pair of gapless helical edge states. Here, we utilize the recently discovered altermagnetism to circumvent this fundamental constraint. We demonstrate the realization of a unique QSH phase possessing multiple pairs of gapless helical edge states in altermagnetic multilayers. This exotic QSH phase, characterized by a mirror-spin Chern number, emerges from the interplay of spin-orbit coupling and $d$-wave altermagnetic ordering. Moreover, using first-principles calculations, we identify altermagnetic Fe$_2$Se$_2$O multilayers as promising material candidates, in which the number of gapless helical edge states scales linearly with the number of layers, leading to a correspondingly large, exactly quantized, and experimentally accessible spin-Hall conductance. Our findings unveil a new mechanism for stabilizing  multiple pairs of gapless helical edge states, significantly expanding the scope of QSH effects, and provide a blueprint for utilizing altermagnetism to engineer desired topological phases.

\end{abstract}


%Use showkeys class option if keyword

\maketitle

\textit{\textcolor{blue}{Introduction. ---}} 
In recent decades, topological phases of matter have garnered considerable attention in the field of condensed-matter physics~\cite{RevModPhys.82.3045,RevModPhys.83.1057,RevModPhys.88.021004,RevModPhys.93.025002,RevModPhys.88.035005}. As prominent examples of such phases, the quantum anomalous Hall (QAH) effect~\cite{PhysRevLett.61.2015,PhysRevB.82.184516,science.1187485,Chang167,RevModPhys.95.011002,annurev011417,annurev054144} and the quantum spin Hall (QSH) effect~\cite{PhysRevLett.95.146802,PhysRevLett.95.226801,PhysRevLett.96.106802,science.1133734,science.1148047,annurev140538,adma.202008029} are of particular importance, serving as foundational concepts in the infancy of topological materials. Because of their dissipationless edge state currents, the QAH and QSH effects provide great potential for enabling next-generation low-power-consumption electronic and spintronic devices.  
%The topological invariant is essential for understanding the physics of nontrivial band topology. 
The QAH effect is characterized by an integer Chern number $\mathcal{C}$, which is directly proportional to the number of chiral gapless edge states, in accordance with the bulk-boundary correspondence~\cite{PhysRevLett.49.405,PhysRevLett.61.2015,PhysRevB.82.184516,PhysRevB.74.045125}.
%Due to the bulk-boundary correspondence, the topological invariant is the key quality for capturing the essential physics arising from nontrivial band topology. The QAH effect is characterized by an integer Chern number $\mathcal{C}$, which is directly proportional to the number of chiral gapless edge states connecting the valence and conduction bands~\cite{}.  
%This leads to a quantized Hall-conductance as $\mathcal{C}\frac{e^2}{h}$~\cite{}. 
Compared to QAH systems that are classified by a definite topological invariant such as Chern number $\mathcal{C}$, how to describe the topological nature of the QSH effect is somewhat less straightforward~\cite{PhysRevLett.107.066602,Sheng_2013,srep43049,PhysRevB.80.125327,PhysRevLett.97.036808}. 

For a time-reversal ($\mathcal{T}$) invariant QSH system, it has been shown that the $\mathbb{Z}_2$ index and spin-Chern number ($\mathcal{C}_s$) can both be employed to characterize the nontrivial topology of QSH effects, yielding equivalent descriptions~\cite{PhysRevLett.97.036808,PhysRevB.80.125327}. 
%The $\mathbb{Z}_2$ index refers to a pair of $\mathcal{T}$-symmetry protected gapless helical edge states. 
%indicating the bulk-boundary correspondence in a two-dimensional (2D) $\mathbb{Z}_2$ topological insulator. 
If a $\mathcal{T}$-symmetry breaking perturbation is introduced, which causes the originally gapless edge states to be gapped, the QSH phase can still be characterized by $\mathcal{C}_s$. This topological phase is referred to as the $\mathcal{T}$-broken QSH effect~\cite{PhysRevLett.107.066602,Sheng_2013,srep43049}. Although the spin-Chern number $\mathcal{C}_s$ can remain conserved in the presence of spin nonconserving perturbation~\cite{PhysRevB.80.125327,Sheng_2013,srep43049,srep03435,PhysRevLett.97.036808,PhysRevB.109.155143,PhysRevB.110.035161,PhysRevB.110.L161104}, it does not correspond to any gapless edge state connecting the valence and conduction bands in $\mathcal{T}$-broken QSH insulators. Moreover, this breakdown of bulk-boundary correspondence also occurs in well-established $\mathcal{T}$-invariant QSH systems since $\mathcal{T}$-symmetry permits only a $\mathbb{Z}_2$-classified QSH effect that possesses one pair of helical edge states~\cite{PhysRevB.109.155143,PhysRevB.110.035161,Zhou_2015,PhysRevB.91.235451,PhysRevB.110.L161104}. 

\begin{figure}
    \centering
    \includegraphics[width=1.00\linewidth]{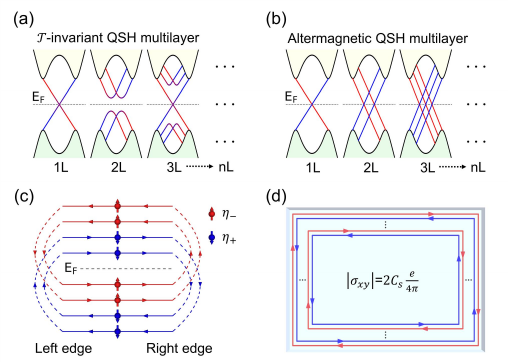}
    \caption{Schematic illustration of QSH effects in $\mathcal{T}$-invariant and altermagnetic multilayers. The characteristics of spin-resolved edge states in (a) $\mathcal{T}$-invariant and (b) altermagnetic multilayers. (c) In the spin transport diagram of altermagnetic multilayers, electronic states with the with spin polarization are connected through chiral edge channels, forming closed-loop conduction paths. (d) Multiple helical edge state pairs in altermagnetic multilayers allow the existence of exactly quantized SHC.
    \label{Fig.1}}
\end{figure}

%To intuitively illustrate the violation of $\mathcal{C}_s$-related bulk-boundary correspondence in $\mathcal{T}$-invariant systems, 
To bypass the fundamental constraint in $\mathcal{T}$-invariant systems, we begin by revisiting the topological evolution of nonmagnetic QSH multilayers as the number of stacked layers increases~\cite{PhysRevB.109.155143,Zhou_2015,PhysRevB.91.235451,PhysRevLett.107.256801,Pan_2014,PhysRevB.82.195438}, 
%, where each individual layer is a QSH insulator monolayer carrying $\mathbb{Z}_2=1$ and $\mathcal{C}_s=1$,  
as depicted in Fig.~\ref{Fig.1}(a).  Each individual layer carries $\mathbb{Z}_2=1$ and $\mathcal{C}_s=1$. 
Here, we assume that the interlayer coupling can be treated as a perturbation. This implies that $\mathcal{C}_s$ of each QSH layer remains invariant, and consequently, the total $\mathcal{C}_s$ of the multilayer system is proportional to the number of layers. Meanwhile, the $\mathbb{Z}_2$ index displays the oscillated behavior between 0 and 1 due to the interplay of interlayer coupling and interband scattering. According to $\mathbb{Z}_2$ classification, there are two distinct scenarios: one with an even number of layers exhibits a global band gap (i.e., $\mathbb{Z}_2 = 0$) and the other with an odd number of layers hosts one pair of gapless helical edge states (i.e., $\mathbb{Z}_2 = 1$). Therefore, except for the QSH phase with $\mathcal{C}_s=1$, a $\mathcal{T}$-invariant QSH system with higher spin-Chern numbers (i.e., $\mathcal{C}_s>1$) typically lacks multiple pairs of gapless helical edge states~\cite{Zhou_2015,PhysRevB.91.235451,PhysRevB.109.155143,NC10.1038}, hindering the realization of high performance in QSH-based devices with multiple dissipationless edge channels. 
%challenging the bulk-boundary correspondence of $\mathcal{C}_s$ established as a topological invariant~\cite{PhysRevB.80.125327,Zhou_2015,PhysRevB.91.235451}.
%Previously, there have been many efforts to understand the $\mathcal{C}_s$-related bulk-boundary correspondence, such as the feature spectrum topology approach~\cite{PhysRevB.80.125327,PhysRevLett.97.036808,PhysRevB.109.155143,NC10.1038}, but the edge states in such a feature spectrum do not correspond to any current in realistic scenarios, hindering the realization of high performance in QSH-based devices with multiple dissipationless edge channels. 
Therefore, exploring novel mechanisms to ensure the robustness of multiple pairs of gapless helical edge states in QSH systems with extended topologically protected features is highly desirable. 

Here, we utilize the recently discovered altermagnetism to extend the conventional $\mathbb{Z}_2$-classification, achieving the desired QSH phases featuring robust multiple pairs of gapless helical edge states.
Note that altermagnetism is the subject of intense current studies, as it uniquely combines fully symmetry-compensated magnetic moments with non-relativistic spin splitting, giving rise to lots of intriguing phenomena~\cite{PhysRevX.12.040002,PhysRevX.12.040501,Observation07023w,Alter10.1038,AHC00866z,sciadv.aaz8809,PhysRevLett.134.096703,PhysRevLett.134.106801,PhysRevLett.132.036702,PhysRevLett.132.056701,adfm.202409327,NRM10.1038,zhang2025,feng2025}. In altermagnets, spin splitting depends on crystalline symmetry rather than spin–orbit coupling (SOC). As a typical representative, the combination of fourfold rotation with $\mathcal{T}$-symmetry ($C_{4z}\mathcal{T}$) enables $d$-wave altermagnetic ordering, leading to opposite spin-polarized bands connecting orthogonal eigenstates of the horizontal mirror reflection~\cite{zhang2025,feng2025,PhysRevB.111.085127}. Strikingly, if this symmetry endows nontrivial band topology, giving rise to equivalence of spin-Chern numbers and mirror-Chern numbers, we can realize symmetry-protected multiple gapless helical edge states in engineered altermagnetic multilayers, as illustrated in Figs.~\ref{Fig.1}(b) and~\ref{Fig.1}(c). 
%Strikingly, if such crystalline symmetry endows nontrivial band topology, it is expected that a topological crystalline insulator with symmetry-protected spin-polarized edge states would be present~\cite{feng2025}. Therefore, when an altermagnetic monolayer evolves into a crystalline symmetry-protected QSH phase, we can naturally realize engineered gapless helical edge states in altermagnetic multilayer systems, as illustrated in Figs.~\ref{Fig.1}(b) and~\ref{Fig.1}(c). 
%In the following, we firstly employ symmetry arguments and effective model analysis to 
Moreover, the exactly quantized spin-Hall conductance (SHC) linearly increases with the number of layers and is proportional to the spin-Chern number $\mathcal{C}_s$, as shown in Fig.~\ref{Fig.1}(d).  
%linearly enlarges with the increase of layers, as shown in Fig.~\ref{Fig.1}(d). 
%we multiple gapless helical edge states present in van der Waals altermagnetic building blocks, where the quantized spin-Hall conductance (SHC) linearly enlarges with the increase of layers, as shown in Fig.~\ref{Fig.1}(d). 
As concrete demonstrations, we verify by first-principles calculations that the QSH effects with robust edge features can be realized in altermagnetic Fe$_2$Se$_2$O multilayers.

\textit{\textcolor{blue}{General scheme description and model.---}}
To follow this proposal, we consider an altermagnetic system stacked by tetragonal lattices [see Fig.~\ref{Fig.2}(a)]. Because of weak van der Waals interlayer coupling, we only retain the nearest-neighbor interlayer coupling between sublattice $A$ of the lower layer and sublattice $B$ of the upper layer, yet supporting the desired topological states. In this case, the multilayer tight-binding Hamiltonian can be written as
\begin{equation}\label{Eq1}
\begin{split}
H=&t_1\sum_{\langle i,j \rangle}c_{i}^\dagger c_{j}+i\lambda\sum_{\langle i,j\rangle}\nu_{ij}s_zc_{i}^\dagger c_{j} +\sum_{\langle\langle i,j\rangle\rangle}t_m c_{i}^\dagger c_{j} \\
&+\sum_i m_{\mu} s_z c_{i}^\dagger c_{i}+\sum_{i,j} t_\perp c_{i}^\dagger c_{j}+h.c.,
\end{split}
\end{equation}
where $c_{i}^\dagger(c_{i})$ is the creation (annihilation) operator of an electron at site $i$. As schematically demonstrated in Fig.~\ref{Fig.2}(a), the first two terms describe the nearest-neighbor hopping and intrinsic spin-orbit coupling involving the nearest-neighbor bonds with $\nu_{ij}=(\mathbf{d}_{ij}^{1}\times\mathbf{d}_{ij}^{2})_{z}=\pm1$. The third term represents the unconventional $d$-wave interaction arising from hopping between next-nearest neighboring sublattices, where the hopping amplitudes alternate in sign depending on the direction. This is expressed as $t_m^x = -t_m^y = (-1)^{\mu}t_m$, with $\mu = 1, 2$ denoting the sublattice index. The fourth term represents the AFM exchange fields on sublattices with $m_{\mu}=(-1)^{\mu}m$. The final term represents the interlayer coupling between any two adjacent layers.
%monolayer, trilayer, and tetralayer
\begin{figure*}
    \centering
    \includegraphics[width=0.9\linewidth]{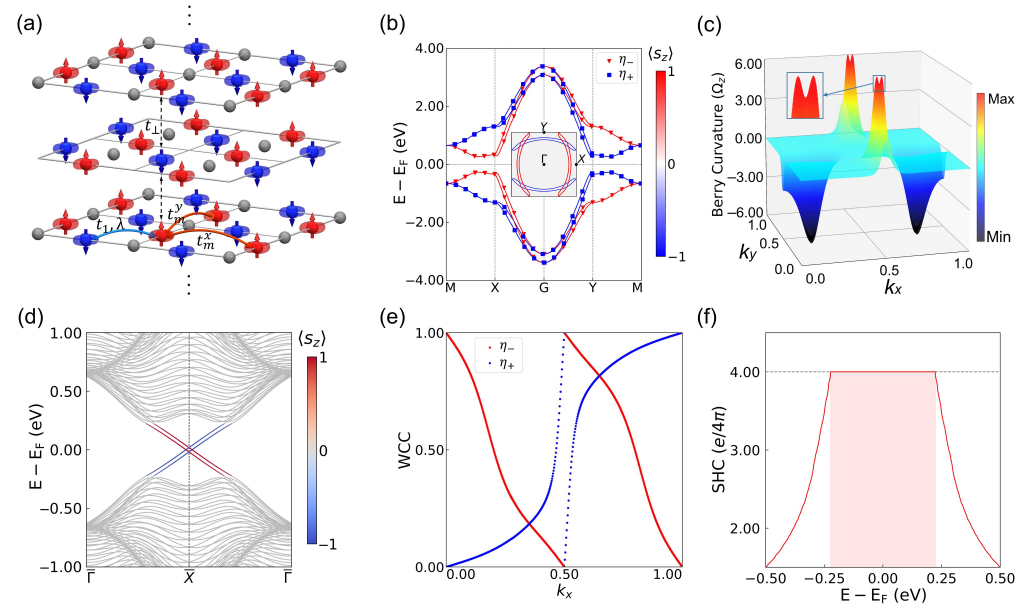}
    \caption{Results of altermagnetic multilayer model Eq.~\ref{Eq1}. 
    (a) Hopping scheme for altermagnetic stacked multilayer tetragonal lattices. The crystalline environment is provided by the nonmagnetic sites (gray atoms). $t_1$, $t_m^{x/y}$, $\lambda$, and $t_{\perp}$ describe the nearest hopping, anisotropic next nearest neighbor hopping, intrinsic SOC interaction, and interlayer coupling strength, respectively. 
    (b) The spin-resolved band structure in the presence of SOC. Inset shows the constant energy contour of the band, corresponding to an energy of -1.0 eV. The red triangles and blue squares are used to respectively denote the states with $\pm$ signs of the eigenvalues of mirror symmetry. 
    (c) The distribution of Berry curvature in the first Brillouin zone. 
    (d) The spectrum of a semi-infinite nanoribbon with (110) edges. 
    (e) The evolution of the Wannier charge centers obtained in the two $M_z$ mirror subspaces, respectively. 
    (f) The exactly quantized spin Hall conductivity $\sigma_{xy}$ inside the bulk gap. The results shown here were obtained with parameters $t_1=0.80$ eV, $t_m=0.20$ eV, $\lambda=0.10$ eV, $t_{\perp}=0.15$ eV, and $m=0.50$ eV.
    \label{Fig.2}}
\end{figure*}
As a concrete example, we focus on the results of an altermagnetic bilayer system, and the general $n$-layered Hamiltonian also is presented in the Supplemental Material (SM)~\cite{SM}. Under the spinful orbital basis $\{ \vert \varphi^A\uparrow\rangle, \vert \varphi^A\downarrow\rangle, \vert \varphi^B\uparrow\rangle, \vert \varphi^B\downarrow\rangle \}$, the bilayer Hamiltonian in momentum space can be written as
%\begin{equation}
%\begin{split}
%H(\boldsymbol{k})&=\left[m-2t_{J}(\cos k_{x}+\cos k_{y})\right]\tau_{0}\sigma_{z}s_{z}\\&+(t_m^y-t_m^x)(\cos k_x-\cos k_y)\tau_0\sigma_zs_0\\&+4t_1\cos(\frac{k_x}{2})\cos(\frac{k_y}{2})\tau_0\sigma_xs_0\\&+4\lambda\sin(\frac{k_{x}}{2})\sin(\frac{k_{y}}{2})\tau_{0}\sigma_{y}s_{z}+t_{z}\tau_{x}\sigma_{x}s_{0}.
%\end{split}
%\end{equation}
\begin{equation}
\begin{split}
H(\boldsymbol{k})&=4t_1\cos(\frac{k_x}{2})\cos(\frac{k_y}{2})\tau_0\sigma_xs_0\\&+4\lambda\sin(\frac{k_{x}}{2})\sin(\frac{k_{y}}{2})\tau_{0}\sigma_{y}s_{z}\\
&+(t_m^x-t_m^y)(\cos k_x-\cos k_y)\tau_0\sigma_zs_0\\&
+m\tau_{0}\sigma_{z}s_{z}+t_{z}\tau_{x}\sigma_{x}s_{0}.
\end{split}
\end{equation}

Here, the $s_\alpha$, $\sigma_\alpha$, and $\tau_\alpha$ ($\alpha = x,y,z$) denote the Pauli matrices acting on the spin, sublattice, and interlayer degrees of freedom, respectively. Once the condition $2|t_m^x-t_m^y| > |m|$ is satisfied, the band inversion occurs. In the absence of SOC, the bilayer system exhibits a topological Weyl semimetal phase characterized by four pairs of spin-polarized Weyl nodes~\cite{SM}.
When SOC is introduced, it is capable of inducing a sizable band gap at these Weyl points, as presented in Fig.~\ref{Fig.2}(b). We can see that the spin-resolved Fermi surface exhibits a clear $d$-wave altermagnetic characteristic protected by the $C_{4z}\mathcal{T}$ symmetry [see inset of Fig.~\ref{Fig.2}(b)].
Meanwhile, the symmetry-enforced Berry curvature shows twin peaks with equal magnitude but opposite sign at the boundaries of $k_x = \pm\pi$ and $k_y = \pm\pi$ [Fig.~\ref{Fig.2}(c)], leading the total integral of the Berry curvature to be zero.
In this case, the spin-resolved topology can be quantified for each spin region by the spin Chern number $\mathcal{C}_s=(\mathcal{C}_{\uparrow}-\mathcal{C}_{\downarrow})/2$, where $\mathcal{C}_{\uparrow\downarrow}$ is obtained by integrating Berry curvatures of all occupied bands of each spin sector separately. The calculated Chern numbers of two spin sectors are $\mathcal{C}_{\uparrow\downarrow}=\pm2$, yielding a high spin Chern number $\mathcal{C}_s=2$. 
This nontrivial topological invariant implies that this system hosts two pairs of helical edge states carrying opposite spin polarization inside the bulk band gap.
To directly illustrate this, we calculate the spectrum of a semi-infinite nanoribbon with (110) edges in Fig.~\ref{Fig.2}(d), which further confirms that the system robustly exists with two pairs of perfect $S_z$ spin-polarized helical edge states despite the $\mathcal{T}$-symmetry breaking, accurately reflecting $\mathcal{C}_s$-related bulk-boundary correspondence.

This remarkable feature can be protected by the $C_{4z}\mathcal{T}$-connected mirror spin coupling inherent to $d$-wave altermagnetic multilayers. A horizontal reflection symmetry $\sigma_h$ (including mirror or glide mirror symmetry) protects crossing points of edge bands with opposite chirality in the invariant plane. For spinful systems, the horizontal reflection operator satisfies $\sigma_h^2 = -1$, enabling Bloch states on this plane to be labeled by their eigenvalue signs $\eta = \pm i$ under the reflection operation, as illustrated in Fig.~\ref{Fig.2}(b). Since any hybridization between $\eta_{+}$ and $\eta_{-}$ bands would break the reflection plane, these crossing points between them at any momentum on the invariant plane are protected. In consequence, the interlayer coupling can only shift the crossing points of edge bands but cannot induce a gap. 
Furthermore, the $C_{4z}\mathcal{T}$ symmetry inverts the eigenvalue of $\sigma_h$, meaning that electronic states related by this symmetry exhibit opposite both spin and $\sigma_h$ eigenvalues. This results in a unique mirror spin coupling, characterized by the commutation relation $[\sigma_h, S_{z}] = 0$ and the anticommutation relations $\{\sigma_h, S_{x/y}\} = 0$ in $d$-wave altermagnetic systems.
%mirror reflection symmetry partitions the system into decoupled subspaces of opposite spin polarization [denoted by red triangles and blue squares in Fig.~\ref{Fig.2}(b)], 
%This symmetry reverses the spin component orthogonal to the mirror plane, while the parallel component remains invariant. As a result, the commutation and anticommutation relations hold, i.e., $[\mathcal{M}_z, S_{z}] = 0$ and $\{\mathcal{M}_z, S_{x/y}\} = 0$. Considering $| \varphi_{k,\eta} \rangle$ as a Bloch state with mirror eigenvalue $\eta$, the expectation values of the spin components yield
%\begin{equation}
%\begin{split}
%\langle S_{z}\rangle&=\langle\varphi_{k,\eta}| \mathcal{M}_z^{-1}\mathcal{M}_z S_{z} \mathcal{M}_z^{-1}\mathcal{M}_z|\varphi_{k,\eta}\rangle \\
%&=|\eta|^{2}\langle S_{z}\rangle=\langle S_{z}\rangle, \\
%\langle S_{x}\rangle&=\langle\varphi_{k,\eta}| \mathcal{M}_z^{-1}\mathcal{M}_z S_{x} \mathcal{M}_z^{-1}\mathcal{M}_z|\varphi_{k,\eta}\rangle \\
%&=-|\eta|^{2}\langle S_{x}\rangle=-\langle S_{x}\rangle=0.
%\end{split}
%\end{equation}
%Similarly, $\langle S_y \rangle = 0$ for all Bloch states. 
Hence, the spin texture of the eigenstates of the system must be polarized along the $z$ direction in the whole 2D Brillouin zone, except at degeneracies. Under the protection of mirror spin coupling, the SOC effect acts only within mirror subspaces, resulting in the mirror-spin Chern insulator characterized by the mirror and spin Chern numbers simultaneously. In analogy to the spin Chern number, the mirror Chern number can be defined as $\mathcal{C}_m=(\mathcal{C}_{+}-\mathcal{C}_{-})/2$, where $\mathcal{C}_{\pm}$ can be obtained in each mirror subspace. In our system, $\mathcal{C}_m$ is also calculated to be 2, which is further confirmed by the Wilson loop calculations performed on the two mirror subspaces, as depicted in Fig.~\ref{Fig.2}(e). Because opposite chiral edge states reside in $C_{4z}\mathcal{T}$-connected mirror subspaces, they are orthogonal to each other and do not hybridize, thereby retaining spin as a good quantum number despite the presence of SOC. It is worth noting that these helical edge states with well-conserved $S_z$ in the altermagnetic bilayer system can contribute an exactly quantized SHC plateau with an amplitude of $|\sigma_{xy}|=4 e/4\pi$ inside the bulk gap, as shown in Fig.~\ref{Fig.2}(f). Moreover, the higher SHC plateau can be obtained by building altermagnetic blocks, thus the novel mirror spin coupling phenomenon in altermagnets opens new avenues for designing multi-channel dissipationless spin transport devices.

%This unique behavior arises from the distinct mirror spin coupling inherent to altermagnetic multilayers. In this altermagnetism, opposite-spin sublattices are connected by rotation or mirror operations but are not connected by translation or inversion~\cite{}.
%Consequently, spin-$\uparrow$ ($\downarrow$) states exclusively occupy the $\eta = -i$ ($+i$) subspace, preventing spin mixing despite strong spin-orbit coupling.

\textit{\textcolor{blue}{Material realization of altermangetic QSH multilayers. ---}} Next, we demonstrate that the emergence of the novel topological characters established by the above tight-binding model can be realized in realistic multilayer materials. To identify potential candidate materials, we performed first-principles calculations within the framework of density-functional theory~\cite{pub.1060429813,pub.1060431417}, and calculation details are included in the SM~\cite{SM}. Here, we demonstrate that the Fe$_2$Se$_2$O multilayers are the promising candidates with multiple dissipationless conducting channels. 
%The layered Fe$_2$Se$_2$O crystallizes in a tetragonal structure with space group P4/mmm (No. 123). 
It is a homologue of experimentally synthesized layered V$_2$Te$_2$O compounds with room-temperature altermagnetic order~\cite{Crystal02864}. We focus mainly on the electronic properties and band topology of the Fe$_2$Se$_2$O bilayer and trilayer, which can be considered as a representation of altermagnetic topological insulator films.

\begin{figure}
    \centering
     \includegraphics[width=1.00\linewidth]{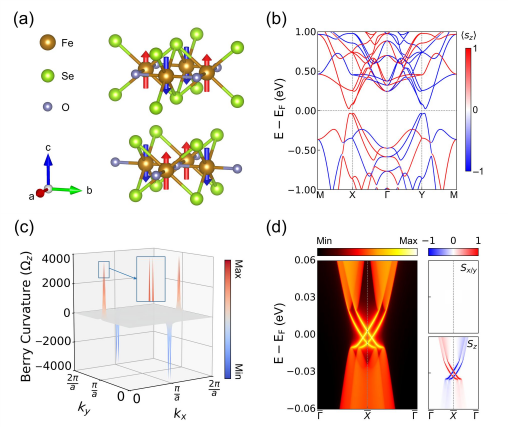}
    \caption{(a) Crystal and magnetic structures of the Fe$_2$Se$_2$O bilayer. The arrows indicate the directions of the magnetic moments of the Fe atoms.
        (b) The spin-resolved band structure of the Fe$_2$Se$_2$O bilayer in the presence of SOC.
        (c) The distribution of the Berry curvature in $k_x$–$k_y$ plane.
        (d) The helical edge states of the Fe$_2$Se$_2$O bilayer on the (110) surface, including spin-resolved edge states projected onto the $S_x$, $S_y$, and $S_z$ components. 
    \label{Fig.3}}
\end{figure}

Figure~\ref{Fig.3}(a) shows the bilayer crystal structure and magnetic configuration with the highest thermodynamic stability~\cite{SM}. This stacking structure is obtained by shifting the upper layer ($\bm{a_1}/2, \bm{a_2}/2$) along the diagonal direction. By the way, although this fractional translation results in films with odd and even layers exhibiting symmorphic and nonsymmorphic mirror symmetry respectively, the nontrivial topology is robust against stacking configurations~\cite{SM}.
In the absence of SOC, the Fe$_2$Se$_2$O bilayer is a Weyl semimetal with spin-polarized Weyl points~\cite{SM}. Once the SOC effect is introduced, the band gaps open immediately at the Weyl points. As illustrated in Fig.~\ref{Fig.3}(b), we plot the spin-resolved electronic band structure. As anticipated, one can clearly see that the electronic structure is strictly separated into spin-up and spin-down states and possesses the same inverted band feature at the X and Y valleys. The nontrivial band topology, associated with $C_{4z}\mathcal{T}$ symmetry, leads to twin Berry curvature peaks of opposite signs that diverge at the X and Y valleys [Fig.~\ref{Fig.3}(c)]. This further gives rise to a high spin Chern number $C_s = 2$. To gain insight into the topologically nontrivial feature, we calculated the edge states of the Fe$_2$Se$_2$O bilayer nanoribbon. As shown in Fig.~\ref{Fig.3}(d), two pairs of gapless helical edge states emerge within the nontrivial gap, in direct agreement with the calculated values of $C_s=2$. Notably, preserved by the $C_{4z}\mathcal{T}$-connected mirror spin coupling, these edge bands exhibit perfectly $S_z$ spin-polarized behavior, and interlayer coupling can only shift the edge band crossing point along the high-symmetry line, but cannot open a gap.

\begin{figure}
    \centering
     \includegraphics[width=1.00\linewidth]{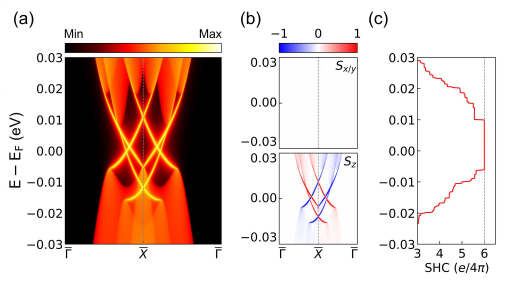}
    \caption{Edge states and spin transport properties of the trilayer Fe$_2$Se$_2$O. 
        (a) The helical edge states on the (110) surface. 
        (b) The spin-resolved helical edge states projected onto the $S_x$, $S_y$, and $S_z$ components. 
        (c) The quantized spin Hall conductivity $\sigma_{xy}$ as a function of energy.
    \label{Fig.4}}
\end{figure}

Expanding our focus to more multiple dissipationless conducting channels, we explore a trilayer structure, highlighting the broad applicability of our design principle by building stacking blocks of van der Waals altermagnetic topological insulators. For the altermagnetic QSH multilayer with high spin Chern numbers, the most exotic signature is the existence of multiple pairs of gapless helical edge states. As presented in Fig.~\ref{Fig.4}(a), there are three pairs of such helical edge states connecting the conduction and valence bands and crossing each other at generic $k$ points. Similar to the bilayer film, these edge bands with opposite mirror eigenvalues are found to cross each other and maintain perfect conservation of the $S_z$ spin component, even in the presence of the SOC effect [Fig.~\ref{Fig.4}(b)]. These $\mathcal{C}_s$-proportional gapless helical edge states in altermagnetic QSH multilayer allow the existence of exactly quantized SHC attributed to the conservation of the spin component $S_z$. To clearly visualize the quantization, Figure~\ref{Fig.4}(c) presents the spin Hall conductivity of the trilayer Fe$_2$Se$_2$O. We can see an exactly quantized SHC plateau with an amplitude of $|\sigma_{xy}|=6e/4\pi$ inside the bulk gap. Thus, these quantized dissipationless spin transport channels can be expanded by increasing the number of van der Waals altermagnetic stacking layers, which is conducive to enhancing the response signals.

\textit{\textcolor{blue}{Summary.---}}
In summary, using an effective tight-binding model and first-principles calculations, we propose an engineering strategy to achieve the altermagnetic QSH phase with multiple robust gapless helical edge states in multilayers.
%crystalline symmetry-protected QSH effect with tunable gapless helical edge states strictly corresponding to the spin Chern number in altermagnetic multilayers.
These robust QSH phases are protected by the $C_{4z}\mathcal{T}$-connected mirror spin coupling inherent to $d$-wave altermagnetic systems.
The first-principles calculations suggest that the altermagnetic Fe$_2$Se$_2$O multilayers are promising candidates, in which the exactly quantized spin-Hall conductance increases linearly with the number of layers, in stark contrast to conventional QSH systems. This mechanism is compatible with van der Waals stacking strategies, making it feasible for experimental realization and device integration.
Our findings not only uncover a novel mechanism for protecting multiple gapless helical edge states in QSH systems with extended topologically protected features, but also offer guidance for engineering altermagnetic topological insulators that enable high-efficiency, multi-channel, and dissipationless spintronic devices.

\textit{\textcolor{blue}{Acknowledgments. ---}}
This work was supported by the National Natural Science Foundation of China (NSFC, Grants No. 12204074, No. 12222402, No. 92365101, No. 12347101, No. 12074108, No. 12447141, and No. 12474151), the Natural Science Foundation of Chongqing (Grants No. 2023NSCQ-JQX0024 and No. CSTB2022NSCQ-MSX0568), the Beijing National Laboratory for Condensed Matter Physics (Grant No. 2024BNLCMPKF025), the Postdoctoral Fellowship Program of CPSF (Grant No. GZC20252254), and the Special Funding for Postdoctoral Research Projects in Chongqing (Grant No. 2024CQBSHTB2036). 

Z.C. and F.Z. contributed equally to this work.

%\bibliographystyle{apsrev4-2}
%\bibliography{ref}


\newpage

\begin{widetext}
\newpage

\setcounter{figure}{0}
\setcounter{equation}{0}
\makeatletter

\makeatother
\renewcommand{\thefigure}{S\arabic{figure}}
\renewcommand{\thetable}{S\Roman{table}}
\renewcommand{\theequation}{S\arabic{equation}}

\begin{center}
	\textbf{
		\large{Supplemental Material for}}
	\vspace{0.2cm}
	
	\textbf{
		\large{
			``Quantum Spin Hall Effect with Extended Topologically Protected Features in Altermangetic Multilayers"}
	}
\end{center}

In this Supplemental Material, we provide details of the effective lattice model Hamiltonian for altermagnetic multilayers, computational methods of first-principles calculations, and the results of the topological states in Fe$_2$Se$_2$O bilayer and trilayer films.

\maketitle

\makeatletter
\def\@hangfrom@section#1#2#3{\@hangfrom{#1#2#3}}
\makeatother

\renewcommand{\thefigure}{S\arabic{figure}}
\renewcommand{\thetable}{S\Roman{table}}
\renewcommand{\theequation}{S\arabic{equation}}

\section{The effective lattice model in altermagnetic multilayers}

To capture the topological origin of the altermagnetic multilayers, we firstly construct a minimum tight-binding model stacked by tetragonal lattices, as shown in Fig.2(a) in the main text. Two magnetic atoms with antiparallel magnetic moments (red and blue atoms) are located at the edge centers, and a nonmagnetic atom is positioned at the corner of the unit cell. The tight-binding Hamiltonian for the monolayer can be written as
\begin{equation}\label{Eq1}
H=t_{1}\sum_{\langle i,j\rangle}c_{i}^{\dagger}c_{j}+i\lambda\sum_{\langle i,j\rangle}\nu_{ij}s_{z}c_{i}^{\dagger}c_{j}+\sum_{\langle(i,j)\rangle}t_{m}c_{i}^{\dagger}c_{j}+\sum_{i}m_{\mu}s_{z}c_{i}^{\dagger}c_{i}+h.c.,
\end{equation}
where $c_{i}^\dagger(c_{i})$ is the creation (annihilation) operator of an electron at site $i$. The first two terms describe the nearest-neighbor hopping and intrinsic spin-orbit coupling involving the nearest-neighbor bonds with $\nu_{ij}=(\mathbf{d}_{ij}^{1}\times\mathbf{d}_{ij}^{2})_{z}=\pm1$. The third term represents the unconventional $d$-wave interaction arising from hopping between next-nearest neighboring sublattices, where the hopping amplitudes alternate in sign depending on the direction. This is expressed as $t_m^x = -t_m^y = (-1)^{\mu}t_m$, with $\mu = 1, 2$ denoting the sublattice index. The fourth term represents the AFM exchange fields on sublattices with $m_{\mu}=(-1)^{\mu}m$. Under the spinful orbital basis $\{ \vert \varphi^A\uparrow\rangle, \vert \varphi^A\downarrow\rangle, \vert \varphi^B\uparrow\rangle, \vert \varphi^B\downarrow\rangle \}$, the monolayer Hamiltonian in momentum space can be written as
\begin{equation}\label{Eq2}
\begin{split}
H(\boldsymbol{k})=4t_{1}\cos(\frac{k_{x}}{2})\cos(\frac{k_{y}}{2})\sigma_{x}s_{0}+4\lambda\sin(\frac{k_{x}}{2})\sin(\frac{k_{y}}{2})\sigma_{y}s_{z}+(t_{m}^{x}-t_{m}^{y})(\cos k_{x}-\cos k_{y})\sigma_{z}s_{0}+m\sigma_{z}s_{z}.
\end{split}
\end{equation}
The $\sigma_\alpha$ and $s_\alpha$ ($\alpha = x, y, z$) denote the Pauli matrices acting on the sublattice and spin degrees of freedom, respectively. Here, we retain only the nearest-neighbor interlayer coupling in the multilayer stacking structure, which connects electrons in sublattice A of the $n$th layer to those in sublattice B of the $(n{-}1)$th and $(n{+}1)$th layers. Accordingly, the coupling between nearest-neighbor layers is given by $H_{\perp}=t_{\perp}\sigma_{x}s_{0}$. Therefore, the Hamiltonian of the altermagnetic model with $n$ layers can be written as
\begin{equation}\label{Eq3}
\begin{split}
H_n\left(\boldsymbol{k}\right)=\begin{pmatrix}H^{(1)}&H_{\perp}&0&0&\cdots&0\\H_{\perp}^\dagger&H^{(2)}&H_{\perp}&0&\cdots&0\\0&H_{\perp}^\dagger&H^{(3)}&H_{\perp}&\cdots&0\\0&0&H_{\perp}^\dagger&H^{(4)}&\cdots&0\\\vdots&\vdots&\vdots&\vdots&\ddots&H_{\perp}\\0&0&0&0&H_{\perp}^\dagger&H^{(n)}\end{pmatrix}.
\end{split}
\end{equation}

\begin{figure}
    \centering
    \includegraphics[scale=1.5]{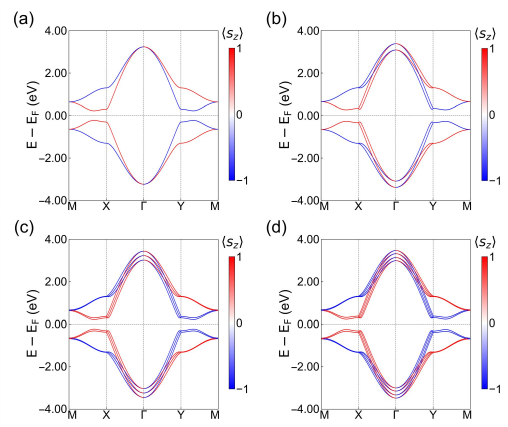}
    \caption{
    Band structures of (a) monolayer, (b) bilayer, (c) trilayer, and (d) four-layer altermagnetic models in the presence of SOC. The parameters are set to \( t_1 = 0.80 \)~eV, $t_m=0.20$ eV, \( \lambda = 0.10 \)~eV, \( t_{\perp} = 0.15 \)~eV, and \( m = 0.50 \)~eV.
    }
   \label{Fig. S1}
\end{figure}

\begin{figure}
    \centering
    \includegraphics[scale=1.5]{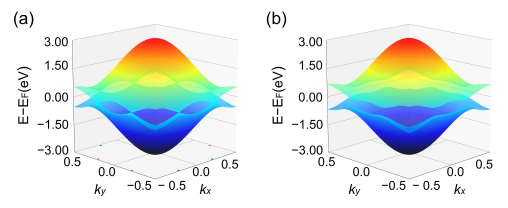}
    \caption{
    3D band structures of the altermagnetic bilayer model in (a) the absence of SOC (\( \lambda = 0 \)) and (b) the presence of SOC (\( \lambda = 0.10 \)~eV). The distribution of Weyl points in the first Brillouin zone is shown in (a). The other parameters are set to \( t_1 = 0.80 \)~eV, $t_m=0.20$ eV,\( t_{\perp} = 0.15 \)~eV, and \( m = 0.50 \)~eV.
    }
   \label{Fig. S2}
\end{figure}

\section{$Ab$ $initio$ computational methods}

First-principles calculations were carried out within the framework of density functional theory (DFT)~\cite{pub.1060429813,pub.1060431417}, as implemented in the Vienna \textit{ab initio} Simulation Package (VASP)~\cite{pub.1060581262,KRESSE199615}.
The exchange-correlation functional was treated within the generalized gradient approximation (GGA) using the Perdew-Burke-Ernzerhof (PBE) scheme~\cite{pub.1060814179}.
The projector-augmented-wave (PAW) method~\cite{pub.1060591374} was employed to describe the electron-ion interaction, and a plane-wave cutoff energy of 500~eV was used in all calculations.
The Brillouin zone was sampled using a $11 \times 11 \times 1$ Monkhorst-Pack $k$-point mesh~\cite{pub.1060521190}.
The convergence thresholds for energy and atomic force were set to $10^{-6}$~eV and 0.01~eV/\AA{}, respectively.
The optimized lattice constants of the Fe$_2$Se$_2$O multilayers are $a_1 = a_2 = 4.05$~\AA{}, and a vacuum layer of 20~\AA was added along the $z$ direction.
The interlayer spacings of the different stacking structures are listed in Table~\ref{Tab. 1}.
To properly describe the weak van der Waals interaction between layers, the Grimme (DFT-D3) method~\cite{jcc.20495} was used.
The PBE+$U_{\text{eff}}$ method with $U_{\text{eff}} = 2.5$~eV was employed to treat the 3$d$ orbitals of Fe atoms~\cite{pub.1024097612}, and the magnetic moment of each Fe atom was initialized to 3.7~$\mu_{\mathrm{B}}$.
We constructed the maximally localized Wannier functions using the WANNIER90 package~\cite{MOSTOFI20142309}, taking into account the Fe $d$ orbitals as well as the $p$ orbitals of Se and O atoms.
We adopted the iterative Green’s function method to calculate the edge states~\cite{MPLopezSancho_1985}, as implemented in the WannierTools package~\cite{WU2018405}.

\section{The topological states in Fe$_2$Se$_2$O bilayer and trilayer films}

\begin{table*}[!htbp]
    \centering
    \renewcommand\arraystretch{1.2}
    \caption{The four stacking structures ($S_{1}$--$S_{4}$) of the Fe$_2$Se$_2$O bilayer. $S_{1}$ corresponds to simple AA stacking. $S_{2}$ and $S_{3}$ are derived by translating the upper layer by $\mathbf{a}_1/2$ and $\mathbf{a}_2/2$ along the $x$ and $y$ axes, respectively. $S_{4}$ is constructed by shifting the upper layer along the diagonal direction by $(\mathbf{a}_1 + \mathbf{a}_2)/2$. Here, $d$ denotes the interlayer distance. $E_b$ represents the binding energy of the Fe$_2$Se$_2$O bilayer for each stacking configuration, which can be obtained by $E_{b}=E_{\text{bilayer}}-2E_{\text{monolayer}}$.}
    \label{Tab1}
    \begin{ruledtabular}
    \begin{tabular}{@{}ccccc@{}}
        & $S_{1}$ & $S_{2}$ & $S_{3}$ & $S_{4}$ \\
        \hline
        $d$ (\AA) & 3.872 & 2.953 & 2.952 & 2.446 \\
        $E_b$ (eV) & -1.224 & -1.336 & -1.336 & -1.416 \\
    \end{tabular}
    \label{Tab. 1}
    \end{ruledtabular}
\end{table*}

\begin{figure}
    \centering
    \includegraphics[scale=1.5]{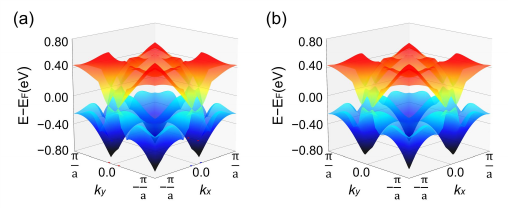}
    \caption{
    The 3D band structures of the Fe$_2$Se$_2$O bilayer in (a) the absence of SOC and (b) the presence of SOC. The distribution of Weyl points in the first Brillouin zone is shown in (a).
    }
   \label{Fig. S3}
\end{figure}

\begin{figure}
    \centering
    \includegraphics[scale=1.5]{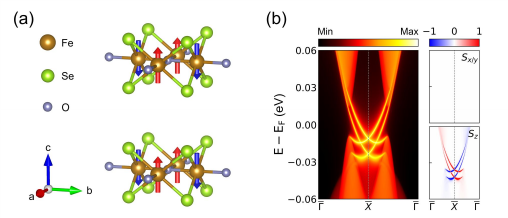}
    \caption{
    (a) Crystal and magnetic structures of the Fe$_2$Se$_2$O bilayer with AA stacking. The arrows indicate the directions of the magnetic moments of the Fe atoms.
    (b) Edge states of the AA-stacked Fe$_2$Se$_2$O bilayer on the (110) surface, including spin-resolved edge states projected onto the $S_x$, $S_y$, and $S_z$ components.
    }
   \label{Fig. S4}
\end{figure}

\begin{figure}
    \centering
    \includegraphics[scale=1.5]{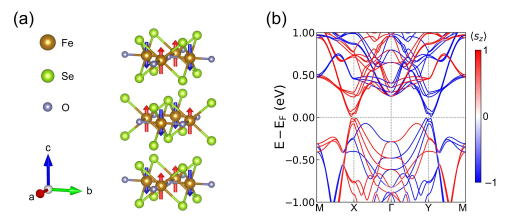}
    \caption{ 
    (a) Crystal and magnetic structures of the Fe$_2$Se$_2$O trilayer. The arrows indicate the directions of the magnetic moments of the Fe atoms.  
    (b) Spin-resolved band structure of the Fe$_2$Se$_2$O trilayer in the presence of SOC.
    }
   \label{Fig. S5}
\end{figure}

~~~\\
~~~\\
~~~\\
~~~\\

\bibliographystyle{apsrev4-2}
%\bibliography{ref-SM}

\begin{thebibliography}{65}%
\makeatletter
\providecommand \@ifxundefined [1]{%
 \@ifx{#1\undefined}
}%
\providecommand \@ifnum [1]{%
 \ifnum #1\expandafter \@firstoftwo
 \else \expandafter \@secondoftwo
 \fi
}%
\providecommand \@ifx [1]{%
 \ifx #1\expandafter \@firstoftwo
 \else \expandafter \@secondoftwo
 \fi
}%
\providecommand \natexlab [1]{#1}%
\providecommand \enquote  [1]{``#1''}%
\providecommand \bibnamefont  [1]{#1}%
\providecommand \bibfnamefont [1]{#1}%
\providecommand \citenamefont [1]{#1}%
\providecommand \href@noop [0]{\@secondoftwo}%
\providecommand \href [0]{\begingroup \@sanitize@url \@href}%
\providecommand \@href[1]{\@@startlink{#1}\@@href}%
\providecommand \@@href[1]{\endgroup#1\@@endlink}%
\providecommand \@sanitize@url [0]{\catcode `\\12\catcode `\$12\catcode `\&12\catcode `\#12\catcode `\^12\catcode `\_12\catcode `\%12\relax}%
\providecommand \@@startlink[1]{}%
\providecommand \@@endlink[0]{}%
\providecommand \url  [0]{\begingroup\@sanitize@url \@url }%
\providecommand \@url [1]{\endgroup\@href {#1}{\urlprefix }}%
\providecommand \urlprefix  [0]{URL }%
\providecommand \Eprint [0]{\href }%
\providecommand \doibase [0]{https://doi.org/}%
\providecommand \selectlanguage [0]{\@gobble}%
\providecommand \bibinfo  [0]{\@secondoftwo}%
\providecommand \bibfield  [0]{\@secondoftwo}%
\providecommand \translation [1]{[#1]}%
\providecommand \BibitemOpen [0]{}%
\providecommand \bibitemStop [0]{}%
\providecommand \bibitemNoStop [0]{.\EOS\space}%
\providecommand \EOS [0]{\spacefactor3000\relax}%
\providecommand \BibitemShut  [1]{\csname bibitem#1\endcsname}%
\let\auto@bib@innerbib\@empty
%</preamble>
\bibitem [{\citenamefont {Hasan}\ and\ \citenamefont {Kane}(2010)}]{RevModPhys.82.3045}%
  \BibitemOpen
  \bibfield  {author} {\bibinfo {author} {\bibfnamefont {M.~Z.}\ \bibnamefont {Hasan}}\ and\ \bibinfo {author} {\bibfnamefont {C.~L.}\ \bibnamefont {Kane}},\ }\bibfield  {title} {\bibinfo {title} {Colloquium: Topological insulators},\ }\href {https://doi.org/10.1103/RevModPhys.82.3045} {\bibfield  {journal} {\bibinfo  {journal} {Rev. Mod. Phys.}\ }\textbf {\bibinfo {volume} {82}},\ \bibinfo {pages} {3045} (\bibinfo {year} {2010})}\BibitemShut {NoStop}%
\bibitem [{\citenamefont {Qi}\ and\ \citenamefont {Zhang}(2011)}]{RevModPhys.83.1057}%
  \BibitemOpen
  \bibfield  {author} {\bibinfo {author} {\bibfnamefont {X.-L.}\ \bibnamefont {Qi}}\ and\ \bibinfo {author} {\bibfnamefont {S.-C.}\ \bibnamefont {Zhang}},\ }\bibfield  {title} {\bibinfo {title} {Topological insulators and superconductors},\ }\href {https://doi.org/10.1103/RevModPhys.83.1057} {\bibfield  {journal} {\bibinfo  {journal} {Rev. Mod. Phys.}\ }\textbf {\bibinfo {volume} {83}},\ \bibinfo {pages} {1057} (\bibinfo {year} {2011})}\BibitemShut {NoStop}%
\bibitem [{\citenamefont {Bansil}\ \emph {et~al.}(2016)\citenamefont {Bansil}, \citenamefont {Lin},\ and\ \citenamefont {Das}}]{RevModPhys.88.021004}%
  \BibitemOpen
  \bibfield  {author} {\bibinfo {author} {\bibfnamefont {A.}~\bibnamefont {Bansil}}, \bibinfo {author} {\bibfnamefont {H.}~\bibnamefont {Lin}},\ and\ \bibinfo {author} {\bibfnamefont {T.}~\bibnamefont {Das}},\ }\bibfield  {title} {\bibinfo {title} {Colloquium: Topological band theory},\ }\href {https://doi.org/10.1103/RevModPhys.88.021004} {\bibfield  {journal} {\bibinfo  {journal} {Rev. Mod. Phys.}\ }\textbf {\bibinfo {volume} {88}},\ \bibinfo {pages} {021004} (\bibinfo {year} {2016})}\BibitemShut {NoStop}%
\bibitem [{\citenamefont {Lv}\ \emph {et~al.}(2021)\citenamefont {Lv}, \citenamefont {Qian},\ and\ \citenamefont {Ding}}]{RevModPhys.93.025002}%
  \BibitemOpen
  \bibfield  {author} {\bibinfo {author} {\bibfnamefont {B.~Q.}\ \bibnamefont {Lv}}, \bibinfo {author} {\bibfnamefont {T.}~\bibnamefont {Qian}},\ and\ \bibinfo {author} {\bibfnamefont {H.}~\bibnamefont {Ding}},\ }\bibfield  {title} {\bibinfo {title} {Experimental perspective on three-dimensional topological semimetals},\ }\href {https://doi.org/10.1103/RevModPhys.93.025002} {\bibfield  {journal} {\bibinfo  {journal} {Rev. Mod. Phys.}\ }\textbf {\bibinfo {volume} {93}},\ \bibinfo {pages} {025002} (\bibinfo {year} {2021})}\BibitemShut {NoStop}%
\bibitem [{\citenamefont {Chiu}\ \emph {et~al.}(2016)\citenamefont {Chiu}, \citenamefont {Teo}, \citenamefont {Schnyder},\ and\ \citenamefont {Ryu}}]{RevModPhys.88.035005}%
  \BibitemOpen
  \bibfield  {author} {\bibinfo {author} {\bibfnamefont {C.-K.}\ \bibnamefont {Chiu}}, \bibinfo {author} {\bibfnamefont {J.~C.~Y.}\ \bibnamefont {Teo}}, \bibinfo {author} {\bibfnamefont {A.~P.}\ \bibnamefont {Schnyder}},\ and\ \bibinfo {author} {\bibfnamefont {S.}~\bibnamefont {Ryu}},\ }\bibfield  {title} {\bibinfo {title} {Classification of topological quantum matter with symmetries},\ }\href {https://doi.org/10.1103/RevModPhys.88.035005} {\bibfield  {journal} {\bibinfo  {journal} {Rev. Mod. Phys.}\ }\textbf {\bibinfo {volume} {88}},\ \bibinfo {pages} {035005} (\bibinfo {year} {2016})}\BibitemShut {NoStop}%
\bibitem [{\citenamefont {Haldane}(1988)}]{PhysRevLett.61.2015}%
  \BibitemOpen
  \bibfield  {author} {\bibinfo {author} {\bibfnamefont {F.~D.~M.}\ \bibnamefont {Haldane}},\ }\bibfield  {title} {\bibinfo {title} {Model for a quantum {Hall} effect without landau levels: Condensed-matter realization of the "parity anomaly"},\ }\href {https://doi.org/10.1103/PhysRevLett.61.2015} {\bibfield  {journal} {\bibinfo  {journal} {Phys. Rev. Lett.}\ }\textbf {\bibinfo {volume} {61}},\ \bibinfo {pages} {2015} (\bibinfo {year} {1988})}\BibitemShut {NoStop}%
\bibitem [{\citenamefont {Qi}\ \emph {et~al.}(2010)\citenamefont {Qi}, \citenamefont {Hughes},\ and\ \citenamefont {Zhang}}]{PhysRevB.82.184516}%
  \BibitemOpen
  \bibfield  {author} {\bibinfo {author} {\bibfnamefont {X.-L.}\ \bibnamefont {Qi}}, \bibinfo {author} {\bibfnamefont {T.~L.}\ \bibnamefont {Hughes}},\ and\ \bibinfo {author} {\bibfnamefont {S.-C.}\ \bibnamefont {Zhang}},\ }\bibfield  {title} {\bibinfo {title} {Chiral topological superconductor from the quantum {Hall} state},\ }\href {https://doi.org/10.1103/PhysRevB.82.184516} {\bibfield  {journal} {\bibinfo  {journal} {Phys. Rev. B}\ }\textbf {\bibinfo {volume} {82}},\ \bibinfo {pages} {184516} (\bibinfo {year} {2010})}\BibitemShut {NoStop}%
\bibitem [{\citenamefont {Yu}\ \emph {et~al.}(2010)\citenamefont {Yu}, \citenamefont {Zhang}, \citenamefont {Zhang}, \citenamefont {Zhang}, \citenamefont {Dai},\ and\ \citenamefont {Fang}}]{science.1187485}%
  \BibitemOpen
  \bibfield  {author} {\bibinfo {author} {\bibfnamefont {R.}~\bibnamefont {Yu}}, \bibinfo {author} {\bibfnamefont {W.}~\bibnamefont {Zhang}}, \bibinfo {author} {\bibfnamefont {H.-J.}\ \bibnamefont {Zhang}}, \bibinfo {author} {\bibfnamefont {S.-C.}\ \bibnamefont {Zhang}}, \bibinfo {author} {\bibfnamefont {X.}~\bibnamefont {Dai}},\ and\ \bibinfo {author} {\bibfnamefont {Z.}~\bibnamefont {Fang}},\ }\bibfield  {title} {\bibinfo {title} {Quantized anomalous {Hall} effect in magnetic topological insulators},\ }\href {https://doi.org/10.1126/science.1187485} {\bibfield  {journal} {\bibinfo  {journal} {Science}\ }\textbf {\bibinfo {volume} {329}},\ \bibinfo {pages} {61} (\bibinfo {year} {2010})}\BibitemShut {NoStop}%
\bibitem [{\citenamefont {Chang}\ \emph {et~al.}(2013)\citenamefont {Chang}, \citenamefont {Zhang}, \citenamefont {Feng} \emph {et~al.}}]{Chang167}%
  \BibitemOpen
  \bibfield  {author} {\bibinfo {author} {\bibfnamefont {C.-Z.}\ \bibnamefont {Chang}}, \bibinfo {author} {\bibfnamefont {J.}~\bibnamefont {Zhang}}, \bibinfo {author} {\bibfnamefont {X.}~\bibnamefont {Feng}}, \emph {et~al.},\ }\bibfield  {title} {\bibinfo {title} {Experimental observation of the quantum anomalous {Hall} effect in a magnetic topological insulator},\ }\href {https://doi.org/10.1126/science.1234414} {\bibfield  {journal} {\bibinfo  {journal} {Science}\ }\textbf {\bibinfo {volume} {340}},\ \bibinfo {pages} {167} (\bibinfo {year} {2013})}\BibitemShut {NoStop}%
\bibitem [{\citenamefont {Chang}\ \emph {et~al.}(2023)\citenamefont {Chang}, \citenamefont {Liu},\ and\ \citenamefont {MacDonald}}]{RevModPhys.95.011002}%
  \BibitemOpen
  \bibfield  {author} {\bibinfo {author} {\bibfnamefont {C.-Z.}\ \bibnamefont {Chang}}, \bibinfo {author} {\bibfnamefont {C.-X.}\ \bibnamefont {Liu}},\ and\ \bibinfo {author} {\bibfnamefont {A.~H.}\ \bibnamefont {MacDonald}},\ }\bibfield  {title} {\bibinfo {title} {Colloquium: Quantum anomalous {Hall} effect},\ }\href {https://doi.org/10.1103/RevModPhys.95.011002} {\bibfield  {journal} {\bibinfo  {journal} {Rev. Mod. Phys.}\ }\textbf {\bibinfo {volume} {95}},\ \bibinfo {pages} {011002} (\bibinfo {year} {2023})}\BibitemShut {NoStop}%
\bibitem [{\citenamefont {Liu}\ \emph {et~al.}(2016)\citenamefont {Liu}, \citenamefont {Zhang},\ and\ \citenamefont {Qi}}]{annurev011417}%
  \BibitemOpen
  \bibfield  {author} {\bibinfo {author} {\bibfnamefont {C.-X.}\ \bibnamefont {Liu}}, \bibinfo {author} {\bibfnamefont {S.-C.}\ \bibnamefont {Zhang}},\ and\ \bibinfo {author} {\bibfnamefont {X.-L.}\ \bibnamefont {Qi}},\ }\bibfield  {title} {\bibinfo {title} {The quantum anomalous {Hall} effect: Theory and experiment},\ }\href {https://doi.org/https://doi.org/10.1146/annurev-conmatphys-031115-011417} {\bibfield  {journal} {\bibinfo  {journal} {Annu. Rev. Condens. Matter Phys.}\ }\textbf {\bibinfo {volume} {7}},\ \bibinfo {pages} {301} (\bibinfo {year} {2016})}\BibitemShut {NoStop}%
\bibitem [{\citenamefont {He}\ \emph {et~al.}(2018)\citenamefont {He}, \citenamefont {Wang},\ and\ \citenamefont {Xue}}]{annurev054144}%
  \BibitemOpen
  \bibfield  {author} {\bibinfo {author} {\bibfnamefont {K.}~\bibnamefont {He}}, \bibinfo {author} {\bibfnamefont {Y.}~\bibnamefont {Wang}},\ and\ \bibinfo {author} {\bibfnamefont {Q.-K.}\ \bibnamefont {Xue}},\ }\bibfield  {title} {\bibinfo {title} {Topological materials: Quantum anomalous {Hall} system},\ }\href {https://doi.org/https://doi.org/10.1146/annurev-conmatphys-033117-054144} {\bibfield  {journal} {\bibinfo  {journal} {Annu. Rev. Condens. Matter Phys.}\ }\textbf {\bibinfo {volume} {9}},\ \bibinfo {pages} {329} (\bibinfo {year} {2018})}\BibitemShut {NoStop}%
\bibitem [{\citenamefont {Kane}\ and\ \citenamefont {Mele}(2005{\natexlab{a}})}]{PhysRevLett.95.146802}%
  \BibitemOpen
  \bibfield  {author} {\bibinfo {author} {\bibfnamefont {C.~L.}\ \bibnamefont {Kane}}\ and\ \bibinfo {author} {\bibfnamefont {E.~J.}\ \bibnamefont {Mele}},\ }\bibfield  {title} {\bibinfo {title} {${Z}_{2}$ topological order and the quantum spin hall effect},\ }\href {https://doi.org/10.1103/PhysRevLett.95.146802} {\bibfield  {journal} {\bibinfo  {journal} {Phys. Rev. Lett.}\ }\textbf {\bibinfo {volume} {95}},\ \bibinfo {pages} {146802} (\bibinfo {year} {2005}{\natexlab{a}})}\BibitemShut {NoStop}%
\bibitem [{\citenamefont {Kane}\ and\ \citenamefont {Mele}(2005{\natexlab{b}})}]{PhysRevLett.95.226801}%
  \BibitemOpen
  \bibfield  {author} {\bibinfo {author} {\bibfnamefont {C.~L.}\ \bibnamefont {Kane}}\ and\ \bibinfo {author} {\bibfnamefont {E.~J.}\ \bibnamefont {Mele}},\ }\bibfield  {title} {\bibinfo {title} {Quantum spin {Hall} effect in graphene},\ }\href {https://doi.org/10.1103/PhysRevLett.95.226801} {\bibfield  {journal} {\bibinfo  {journal} {Phys. Rev. Lett.}\ }\textbf {\bibinfo {volume} {95}},\ \bibinfo {pages} {226801} (\bibinfo {year} {2005}{\natexlab{b}})}\BibitemShut {NoStop}%
\bibitem [{\citenamefont {Bernevig}\ and\ \citenamefont {Zhang}(2006)}]{PhysRevLett.96.106802}%
  \BibitemOpen
  \bibfield  {author} {\bibinfo {author} {\bibfnamefont {B.~A.}\ \bibnamefont {Bernevig}}\ and\ \bibinfo {author} {\bibfnamefont {S.-C.}\ \bibnamefont {Zhang}},\ }\bibfield  {title} {\bibinfo {title} {Quantum spin {Hall} effect},\ }\href {https://doi.org/10.1103/PhysRevLett.96.106802} {\bibfield  {journal} {\bibinfo  {journal} {Phys. Rev. Lett.}\ }\textbf {\bibinfo {volume} {96}},\ \bibinfo {pages} {106802} (\bibinfo {year} {2006})}\BibitemShut {NoStop}%
\bibitem [{\citenamefont {Bernevig}\ \emph {et~al.}(2006)\citenamefont {Bernevig}, \citenamefont {Hughes},\ and\ \citenamefont {Zhang}}]{science.1133734}%
  \BibitemOpen
  \bibfield  {author} {\bibinfo {author} {\bibfnamefont {B.~A.}\ \bibnamefont {Bernevig}}, \bibinfo {author} {\bibfnamefont {T.~L.}\ \bibnamefont {Hughes}},\ and\ \bibinfo {author} {\bibfnamefont {S.-C.}\ \bibnamefont {Zhang}},\ }\bibfield  {title} {\bibinfo {title} {Quantum spin {Hall} effect and topological phase transition in {HgTe} quantum wells},\ }\href {https://doi.org/10.1126/science.1133734} {\bibfield  {journal} {\bibinfo  {journal} {Science}\ }\textbf {\bibinfo {volume} {314}},\ \bibinfo {pages} {1757} (\bibinfo {year} {2006})}\BibitemShut {NoStop}%
\bibitem [{\citenamefont {König}\ \emph {et~al.}(2007)\citenamefont {König}, \citenamefont {Wiedmann}, \citenamefont {Brüne}, \citenamefont {Roth}, \citenamefont {Buhmann}, \citenamefont {Molenkamp}, \citenamefont {Qi},\ and\ \citenamefont {Zhang}}]{science.1148047}%
  \BibitemOpen
  \bibfield  {author} {\bibinfo {author} {\bibfnamefont {M.}~\bibnamefont {König}}, \bibinfo {author} {\bibfnamefont {S.}~\bibnamefont {Wiedmann}}, \bibinfo {author} {\bibfnamefont {C.}~\bibnamefont {Brüne}}, \bibinfo {author} {\bibfnamefont {A.}~\bibnamefont {Roth}}, \bibinfo {author} {\bibfnamefont {H.}~\bibnamefont {Buhmann}}, \bibinfo {author} {\bibfnamefont {L.~W.}\ \bibnamefont {Molenkamp}}, \bibinfo {author} {\bibfnamefont {X.-L.}\ \bibnamefont {Qi}},\ and\ \bibinfo {author} {\bibfnamefont {S.-C.}\ \bibnamefont {Zhang}},\ }\bibfield  {title} {\bibinfo {title} {Quantum spin {Hall} insulator state in {HgTe} quantum wells},\ }\href {https://doi.org/10.1126/science.1148047} {\bibfield  {journal} {\bibinfo  {journal} {Science}\ }\textbf {\bibinfo {volume} {318}},\ \bibinfo {pages} {766} (\bibinfo {year} {2007})}\BibitemShut {NoStop}%
\bibitem [{\citenamefont {Maciejko}\ \emph {et~al.}(2011)\citenamefont {Maciejko}, \citenamefont {Hughes},\ and\ \citenamefont {Zhang}}]{annurev140538}%
  \BibitemOpen
  \bibfield  {author} {\bibinfo {author} {\bibfnamefont {J.}~\bibnamefont {Maciejko}}, \bibinfo {author} {\bibfnamefont {T.~L.}\ \bibnamefont {Hughes}},\ and\ \bibinfo {author} {\bibfnamefont {S.-C.}\ \bibnamefont {Zhang}},\ }\bibfield  {title} {\bibinfo {title} {The quantum spin {Hall} effect},\ }\href {https://doi.org/https://doi.org/10.1146/annurev-conmatphys-062910-140538} {\bibfield  {journal} {\bibinfo  {journal} {Annu. Rev. Condens. Matter Phys.}\ }\textbf {\bibinfo {volume} {2}},\ \bibinfo {pages} {31} (\bibinfo {year} {2011})}\BibitemShut {NoStop}%
\bibitem [{\citenamefont {Lodge}\ \emph {et~al.}(2021)\citenamefont {Lodge}, \citenamefont {Yang}, \citenamefont {Mukherjee},\ and\ \citenamefont {Weber}}]{adma.202008029}%
  \BibitemOpen
  \bibfield  {author} {\bibinfo {author} {\bibfnamefont {M.~S.}\ \bibnamefont {Lodge}}, \bibinfo {author} {\bibfnamefont {S.~A.}\ \bibnamefont {Yang}}, \bibinfo {author} {\bibfnamefont {S.}~\bibnamefont {Mukherjee}},\ and\ \bibinfo {author} {\bibfnamefont {B.}~\bibnamefont {Weber}},\ }\bibfield  {title} {\bibinfo {title} {Atomically thin quantum spin {Hall} insulators},\ }\href {https://doi.org/https://doi.org/10.1002/adma.202008029} {\bibfield  {journal} {\bibinfo  {journal} {Adv. Mater.}\ }\textbf {\bibinfo {volume} {33}},\ \bibinfo {pages} {2008029} (\bibinfo {year} {2021})}\BibitemShut {NoStop}%
\bibitem [{\citenamefont {Thouless}\ \emph {et~al.}(1982)\citenamefont {Thouless}, \citenamefont {Kohmoto}, \citenamefont {Nightingale},\ and\ \citenamefont {den Nijs}}]{PhysRevLett.49.405}%
  \BibitemOpen
  \bibfield  {author} {\bibinfo {author} {\bibfnamefont {D.~J.}\ \bibnamefont {Thouless}}, \bibinfo {author} {\bibfnamefont {M.}~\bibnamefont {Kohmoto}}, \bibinfo {author} {\bibfnamefont {M.~P.}\ \bibnamefont {Nightingale}},\ and\ \bibinfo {author} {\bibfnamefont {M.}~\bibnamefont {den Nijs}},\ }\bibfield  {title} {\bibinfo {title} {Quantized {Hall} conductance in a two-dimensional periodic potential},\ }\href {https://doi.org/10.1103/PhysRevLett.49.405} {\bibfield  {journal} {\bibinfo  {journal} {Phys. Rev. Lett.}\ }\textbf {\bibinfo {volume} {49}},\ \bibinfo {pages} {405} (\bibinfo {year} {1982})}\BibitemShut {NoStop}%
\bibitem [{\citenamefont {Qi}\ \emph {et~al.}(2006)\citenamefont {Qi}, \citenamefont {Wu},\ and\ \citenamefont {Zhang}}]{PhysRevB.74.045125}%
  \BibitemOpen
  \bibfield  {author} {\bibinfo {author} {\bibfnamefont {X.-L.}\ \bibnamefont {Qi}}, \bibinfo {author} {\bibfnamefont {Y.-S.}\ \bibnamefont {Wu}},\ and\ \bibinfo {author} {\bibfnamefont {S.-C.}\ \bibnamefont {Zhang}},\ }\bibfield  {title} {\bibinfo {title} {General theorem relating the bulk topological number to edge states in two-dimensional insulators},\ }\href {https://doi.org/10.1103/PhysRevB.74.045125} {\bibfield  {journal} {\bibinfo  {journal} {Phys. Rev. B}\ }\textbf {\bibinfo {volume} {74}},\ \bibinfo {pages} {045125} (\bibinfo {year} {2006})}\BibitemShut {NoStop}%
\bibitem [{\citenamefont {Yang}\ \emph {et~al.}(2011)\citenamefont {Yang}, \citenamefont {Xu}, \citenamefont {Sheng}, \citenamefont {Wang}, \citenamefont {Xing},\ and\ \citenamefont {Sheng}}]{PhysRevLett.107.066602}%
  \BibitemOpen
  \bibfield  {author} {\bibinfo {author} {\bibfnamefont {Y.}~\bibnamefont {Yang}}, \bibinfo {author} {\bibfnamefont {Z.}~\bibnamefont {Xu}}, \bibinfo {author} {\bibfnamefont {L.}~\bibnamefont {Sheng}}, \bibinfo {author} {\bibfnamefont {B.}~\bibnamefont {Wang}}, \bibinfo {author} {\bibfnamefont {D.~Y.}\ \bibnamefont {Xing}},\ and\ \bibinfo {author} {\bibfnamefont {D.~N.}\ \bibnamefont {Sheng}},\ }\bibfield  {title} {\bibinfo {title} {Time-reversal-symmetry-broken quantum spin {Hall} effect},\ }\href {https://doi.org/10.1103/PhysRevLett.107.066602} {\bibfield  {journal} {\bibinfo  {journal} {Phys. Rev. Lett.}\ }\textbf {\bibinfo {volume} {107}},\ \bibinfo {pages} {066602} (\bibinfo {year} {2011})}\BibitemShut {NoStop}%
\bibitem [{\citenamefont {Sheng}\ \emph {et~al.}(2013)\citenamefont {Sheng}, \citenamefont {Li}, \citenamefont {Yang}, \citenamefont {Sheng},\ and\ \citenamefont {Xing}}]{Sheng_2013}%
  \BibitemOpen
  \bibfield  {author} {\bibinfo {author} {\bibfnamefont {L.}~\bibnamefont {Sheng}}, \bibinfo {author} {\bibfnamefont {H.-C.}\ \bibnamefont {Li}}, \bibinfo {author} {\bibfnamefont {Y.-Y.}\ \bibnamefont {Yang}}, \bibinfo {author} {\bibfnamefont {D.-N.}\ \bibnamefont {Sheng}},\ and\ \bibinfo {author} {\bibfnamefont {D.-Y.}\ \bibnamefont {Xing}},\ }\bibfield  {title} {\bibinfo {title} {Spin {Chern} numbers and time-reversal-symmetry-broken quantum spin {Hall} effect},\ }\href {https://doi.org/10.1088/1674-1056/22/6/067201} {\bibfield  {journal} {\bibinfo  {journal} {Chin. Phys. B}\ }\textbf {\bibinfo {volume} {22}},\ \bibinfo {pages} {067201} (\bibinfo {year} {2013})}\BibitemShut {NoStop}%
\bibitem [{\citenamefont {Luo}\ \emph {et~al.}(2017)\citenamefont {Luo}, \citenamefont {Shao}, \citenamefont {Deng}, \citenamefont {Deng},\ and\ \citenamefont {Sheng}}]{srep43049}%
  \BibitemOpen
  \bibfield  {author} {\bibinfo {author} {\bibfnamefont {W.}~\bibnamefont {Luo}}, \bibinfo {author} {\bibfnamefont {D.~X.}\ \bibnamefont {Shao}}, \bibinfo {author} {\bibfnamefont {M.-X.}\ \bibnamefont {Deng}}, \bibinfo {author} {\bibfnamefont {W.~Y.}\ \bibnamefont {Deng}},\ and\ \bibinfo {author} {\bibfnamefont {L.}~\bibnamefont {Sheng}},\ }\bibfield  {title} {\bibinfo {title} {Time-reversal-breaking induced quantum spin {Hall} effect},\ }\href {https://doi.org/10.1038/srep43049} {\bibfield  {journal} {\bibinfo  {journal} {Sci. Rep.}\ }\textbf {\bibinfo {volume} {7}} (\bibinfo {year} {2017})}\BibitemShut {NoStop}%
\bibitem [{\citenamefont {Prodan}(2009)}]{PhysRevB.80.125327}%
  \BibitemOpen
  \bibfield  {author} {\bibinfo {author} {\bibfnamefont {E.}~\bibnamefont {Prodan}},\ }\bibfield  {title} {\bibinfo {title} {Robustness of the spin-{Chern} number},\ }\href {https://doi.org/10.1103/PhysRevB.80.125327} {\bibfield  {journal} {\bibinfo  {journal} {Phys. Rev. B}\ }\textbf {\bibinfo {volume} {80}},\ \bibinfo {pages} {125327} (\bibinfo {year} {2009})}\BibitemShut {NoStop}%
\bibitem [{\citenamefont {Sheng}\ \emph {et~al.}(2006)\citenamefont {Sheng}, \citenamefont {Weng}, \citenamefont {Sheng},\ and\ \citenamefont {Haldane}}]{PhysRevLett.97.036808}%
  \BibitemOpen
  \bibfield  {author} {\bibinfo {author} {\bibfnamefont {D.~N.}\ \bibnamefont {Sheng}}, \bibinfo {author} {\bibfnamefont {Z.~Y.}\ \bibnamefont {Weng}}, \bibinfo {author} {\bibfnamefont {L.}~\bibnamefont {Sheng}},\ and\ \bibinfo {author} {\bibfnamefont {F.~D.~M.}\ \bibnamefont {Haldane}},\ }\bibfield  {title} {\bibinfo {title} {Quantum spin-{Hall} effect and topologically invariant {Chern} numbers},\ }\href {https://doi.org/10.1103/PhysRevLett.97.036808} {\bibfield  {journal} {\bibinfo  {journal} {Phys. Rev. Lett.}\ }\textbf {\bibinfo {volume} {97}},\ \bibinfo {pages} {036808} (\bibinfo {year} {2006})}\BibitemShut {NoStop}%
\bibitem [{\citenamefont {Ezawa}(2013)}]{srep03435}%
  \BibitemOpen
  \bibfield  {author} {\bibinfo {author} {\bibfnamefont {M.}~\bibnamefont {Ezawa}},\ }\bibfield  {title} {\bibinfo {title} {High spin-{Chern} insulators with magnetic order},\ }\href {https://doi.org/10.1038/srep03435} {\bibfield  {journal} {\bibinfo  {journal} {Sci. Rep.}\ }\textbf {\bibinfo {volume} {3}} (\bibinfo {year} {2013})}\BibitemShut {NoStop}%
\bibitem [{\citenamefont {Yao}\ \emph {et~al.}(2024)\citenamefont {Yao}, \citenamefont {Zhou}, \citenamefont {Hung}, \citenamefont {Lin}, \citenamefont {Bansil},\ and\ \citenamefont {Chang}}]{PhysRevB.109.155143}%
  \BibitemOpen
  \bibfield  {author} {\bibinfo {author} {\bibfnamefont {Y.-T.}\ \bibnamefont {Yao}}, \bibinfo {author} {\bibfnamefont {X.}~\bibnamefont {Zhou}}, \bibinfo {author} {\bibfnamefont {Y.-C.}\ \bibnamefont {Hung}}, \bibinfo {author} {\bibfnamefont {H.}~\bibnamefont {Lin}}, \bibinfo {author} {\bibfnamefont {A.}~\bibnamefont {Bansil}},\ and\ \bibinfo {author} {\bibfnamefont {T.-R.}\ \bibnamefont {Chang}},\ }\bibfield  {title} {\bibinfo {title} {Feature-energy duality of topological boundary states in a multilayer quantum spin {Hall} insulator},\ }\href {https://doi.org/10.1103/PhysRevB.109.155143} {\bibfield  {journal} {\bibinfo  {journal} {Phys. Rev. B}\ }\textbf {\bibinfo {volume} {109}},\ \bibinfo {pages} {155143} (\bibinfo {year} {2024})}\BibitemShut {NoStop}%
\bibitem [{\citenamefont {Liu}\ \emph {et~al.}(2024{\natexlab{a}})\citenamefont {Liu}, \citenamefont {Liu}, \citenamefont {Li}, \citenamefont {Wu},\ and\ \citenamefont {Liu}}]{PhysRevB.110.035161}%
  \BibitemOpen
  \bibfield  {author} {\bibinfo {author} {\bibfnamefont {L.}~\bibnamefont {Liu}}, \bibinfo {author} {\bibfnamefont {Y.}~\bibnamefont {Liu}}, \bibinfo {author} {\bibfnamefont {J.}~\bibnamefont {Li}}, \bibinfo {author} {\bibfnamefont {H.}~\bibnamefont {Wu}},\ and\ \bibinfo {author} {\bibfnamefont {Q.}~\bibnamefont {Liu}},\ }\bibfield  {title} {\bibinfo {title} {Orbital doublet driven even-spin {Chern} insulators},\ }\href {https://doi.org/10.1103/PhysRevB.110.035161} {\bibfield  {journal} {\bibinfo  {journal} {Phys. Rev. B}\ }\textbf {\bibinfo {volume} {110}},\ \bibinfo {pages} {035161} (\bibinfo {year} {2024}{\natexlab{a}})}\BibitemShut {NoStop}%
\bibitem [{\citenamefont {Liu}\ \emph {et~al.}(2024{\natexlab{b}})\citenamefont {Liu}, \citenamefont {Liu}, \citenamefont {Li}, \citenamefont {Wu},\ and\ \citenamefont {Liu}}]{PhysRevB.110.L161104}%
  \BibitemOpen
  \bibfield  {author} {\bibinfo {author} {\bibfnamefont {L.}~\bibnamefont {Liu}}, \bibinfo {author} {\bibfnamefont {Y.}~\bibnamefont {Liu}}, \bibinfo {author} {\bibfnamefont {J.}~\bibnamefont {Li}}, \bibinfo {author} {\bibfnamefont {H.}~\bibnamefont {Wu}},\ and\ \bibinfo {author} {\bibfnamefont {Q.}~\bibnamefont {Liu}},\ }\bibfield  {title} {\bibinfo {title} {Quantum spin {Hall} effect protected by spin {U(1)} quasisymmetry},\ }\href {https://doi.org/10.1103/PhysRevB.110.L161104} {\bibfield  {journal} {\bibinfo  {journal} {Phys. Rev. B}\ }\textbf {\bibinfo {volume} {110}},\ \bibinfo {pages} {L161104} (\bibinfo {year} {2024}{\natexlab{b}})}\BibitemShut {NoStop}%
\bibitem [{\citenamefont {Zhou}\ \emph {et~al.}(2015)\citenamefont {Zhou}, \citenamefont {Feng}, \citenamefont {Liu},\ and\ \citenamefont {Yao}}]{Zhou_2015}%
  \BibitemOpen
  \bibfield  {author} {\bibinfo {author} {\bibfnamefont {J.-J.}\ \bibnamefont {Zhou}}, \bibinfo {author} {\bibfnamefont {W.}~\bibnamefont {Feng}}, \bibinfo {author} {\bibfnamefont {G.-B.}\ \bibnamefont {Liu}},\ and\ \bibinfo {author} {\bibfnamefont {Y.}~\bibnamefont {Yao}},\ }\bibfield  {title} {\bibinfo {title} {Topological edge states in single- and multi-layer $\mathrm{Bi}_{4}\mathrm{Br}_{4}$},\ }\href {https://doi.org/10.1088/1367-2630/17/1/015004} {\bibfield  {journal} {\bibinfo  {journal} {New J. Phys.}\ }\textbf {\bibinfo {volume} {17}},\ \bibinfo {pages} {015004} (\bibinfo {year} {2015})}\BibitemShut {NoStop}%
\bibitem [{\citenamefont {Garc\'{\i}a-Mart\'{\i}nez}\ \emph {et~al.}(2015)\citenamefont {Garc\'{\i}a-Mart\'{\i}nez}, \citenamefont {Lado},\ and\ \citenamefont {Fern\'andez-Rossier}}]{PhysRevB.91.235451}%
  \BibitemOpen
  \bibfield  {author} {\bibinfo {author} {\bibfnamefont {N.~A.}\ \bibnamefont {Garc\'{\i}a-Mart\'{\i}nez}}, \bibinfo {author} {\bibfnamefont {J.~L.}\ \bibnamefont {Lado}},\ and\ \bibinfo {author} {\bibfnamefont {J.}~\bibnamefont {Fern\'andez-Rossier}},\ }\bibfield  {title} {\bibinfo {title} {Quantum spin {Hall} phase in multilayer graphene},\ }\href {https://doi.org/10.1103/PhysRevB.91.235451} {\bibfield  {journal} {\bibinfo  {journal} {Phys. Rev. B}\ }\textbf {\bibinfo {volume} {91}},\ \bibinfo {pages} {235451} (\bibinfo {year} {2015})}\BibitemShut {NoStop}%
\bibitem [{\citenamefont {Qiao}\ \emph {et~al.}(2011)\citenamefont {Qiao}, \citenamefont {Tse}, \citenamefont {Jiang}, \citenamefont {Yao},\ and\ \citenamefont {Niu}}]{PhysRevLett.107.256801}%
  \BibitemOpen
  \bibfield  {author} {\bibinfo {author} {\bibfnamefont {Z.}~\bibnamefont {Qiao}}, \bibinfo {author} {\bibfnamefont {W.-K.}\ \bibnamefont {Tse}}, \bibinfo {author} {\bibfnamefont {H.}~\bibnamefont {Jiang}}, \bibinfo {author} {\bibfnamefont {Y.}~\bibnamefont {Yao}},\ and\ \bibinfo {author} {\bibfnamefont {Q.}~\bibnamefont {Niu}},\ }\bibfield  {title} {\bibinfo {title} {Two-dimensional topological insulator state and topological phase transition in bilayer graphene},\ }\href {https://doi.org/10.1103/PhysRevLett.107.256801} {\bibfield  {journal} {\bibinfo  {journal} {Phys. Rev. Lett.}\ }\textbf {\bibinfo {volume} {107}},\ \bibinfo {pages} {256801} (\bibinfo {year} {2011})}\BibitemShut {NoStop}%
\bibitem [{\citenamefont {Pan}\ \emph {et~al.}(2014)\citenamefont {Pan}, \citenamefont {Li}, \citenamefont {Qiao}, \citenamefont {Liu}, \citenamefont {Yao},\ and\ \citenamefont {Yang}}]{Pan_2014}%
  \BibitemOpen
  \bibfield  {author} {\bibinfo {author} {\bibfnamefont {H.}~\bibnamefont {Pan}}, \bibinfo {author} {\bibfnamefont {X.}~\bibnamefont {Li}}, \bibinfo {author} {\bibfnamefont {Z.}~\bibnamefont {Qiao}}, \bibinfo {author} {\bibfnamefont {C.-C.}\ \bibnamefont {Liu}}, \bibinfo {author} {\bibfnamefont {Y.}~\bibnamefont {Yao}},\ and\ \bibinfo {author} {\bibfnamefont {S.~A.}\ \bibnamefont {Yang}},\ }\bibfield  {title} {\bibinfo {title} {Topological metallic phases in spin–orbit coupled bilayer systems},\ }\href {https://doi.org/10.1088/1367-2630/16/12/123015} {\bibfield  {journal} {\bibinfo  {journal} {New J. Phys.}\ }\textbf {\bibinfo {volume} {16}},\ \bibinfo {pages} {123015} (\bibinfo {year} {2014})}\BibitemShut {NoStop}%
\bibitem [{\citenamefont {Cortijo}\ \emph {et~al.}(2010)\citenamefont {Cortijo}, \citenamefont {Grushin},\ and\ \citenamefont {Vozmediano}}]{PhysRevB.82.195438}%
  \BibitemOpen
  \bibfield  {author} {\bibinfo {author} {\bibfnamefont {A.}~\bibnamefont {Cortijo}}, \bibinfo {author} {\bibfnamefont {A.~G.}\ \bibnamefont {Grushin}},\ and\ \bibinfo {author} {\bibfnamefont {M.~A.~H.}\ \bibnamefont {Vozmediano}},\ }\bibfield  {title} {\bibinfo {title} {Topological insulating phases in monolayer and bilayer graphene: An effective action approach},\ }\href {https://doi.org/10.1103/PhysRevB.82.195438} {\bibfield  {journal} {\bibinfo  {journal} {Phys. Rev. B}\ }\textbf {\bibinfo {volume} {82}},\ \bibinfo {pages} {195438} (\bibinfo {year} {2010})}\BibitemShut {NoStop}%
\bibitem [{\citenamefont {Lin}\ \emph {et~al.}(2024)\citenamefont {Lin}, \citenamefont {Palumbo}, \citenamefont {Guo}, \citenamefont {Hwang}, \citenamefont {Blackburn}, \citenamefont {Shoemaker}, \citenamefont {Mahmood}, \citenamefont {Wang}, \citenamefont {Fiete}, \citenamefont {Wieder},\ and\ \citenamefont {Bradlyn}}]{NC10.1038}%
  \BibitemOpen
  \bibfield  {author} {\bibinfo {author} {\bibfnamefont {K.-S.}\ \bibnamefont {Lin}}, \bibinfo {author} {\bibfnamefont {G.}~\bibnamefont {Palumbo}}, \bibinfo {author} {\bibfnamefont {Z.}~\bibnamefont {Guo}}, \bibinfo {author} {\bibfnamefont {Y.}~\bibnamefont {Hwang}}, \bibinfo {author} {\bibfnamefont {J.}~\bibnamefont {Blackburn}}, \bibinfo {author} {\bibfnamefont {D.~P.}\ \bibnamefont {Shoemaker}}, \bibinfo {author} {\bibfnamefont {F.}~\bibnamefont {Mahmood}}, \bibinfo {author} {\bibfnamefont {Z.}~\bibnamefont {Wang}}, \bibinfo {author} {\bibfnamefont {G.~A.}\ \bibnamefont {Fiete}}, \bibinfo {author} {\bibfnamefont {B.~J.}\ \bibnamefont {Wieder}},\ and\ \bibinfo {author} {\bibfnamefont {B.}~\bibnamefont {Bradlyn}},\ }\bibfield  {title} {\bibinfo {title} {Spin-resolved topology and partial axion angles in three-dimensional insulators},\ }\href {https://doi.org/10.1038/s41467-024-44762-w} {\bibfield  {journal} {\bibinfo  {journal} {Nat. Commun.}\ }\textbf {\bibinfo {volume} {15}} (\bibinfo {year}
  {2024})}\BibitemShut {NoStop}%
\bibitem [{\citenamefont {Mazin}(2022)}]{PhysRevX.12.040002}%
  \BibitemOpen
  \bibfield  {author} {\bibinfo {author} {\bibfnamefont {I.}~\bibnamefont {Mazin}} (\bibinfo {collaboration} {The PRX Editors}),\ }\bibfield  {title} {\bibinfo {title} {Editorial: Altermagnetism---a new punch line of fundamental magnetism},\ }\href {https://doi.org/10.1103/PhysRevX.12.040002} {\bibfield  {journal} {\bibinfo  {journal} {Phys. Rev. X}\ }\textbf {\bibinfo {volume} {12}},\ \bibinfo {pages} {040002} (\bibinfo {year} {2022})}\BibitemShut {NoStop}%
\bibitem [{\citenamefont {\ifmmode~\check{S}\else \v{S}\fi{}mejkal}\ \emph {et~al.}(2022)\citenamefont {\ifmmode~\check{S}\else \v{S}\fi{}mejkal}, \citenamefont {Sinova},\ and\ \citenamefont {Jungwirth}}]{PhysRevX.12.040501}%
  \BibitemOpen
  \bibfield  {author} {\bibinfo {author} {\bibfnamefont {L.}~\bibnamefont {\ifmmode~\check{S}\else \v{S}\fi{}mejkal}}, \bibinfo {author} {\bibfnamefont {J.}~\bibnamefont {Sinova}},\ and\ \bibinfo {author} {\bibfnamefont {T.}~\bibnamefont {Jungwirth}},\ }\bibfield  {title} {\bibinfo {title} {Emerging research landscape of altermagnetism},\ }\href {https://doi.org/10.1103/PhysRevX.12.040501} {\bibfield  {journal} {\bibinfo  {journal} {Phys. Rev. X}\ }\textbf {\bibinfo {volume} {12}},\ \bibinfo {pages} {040501} (\bibinfo {year} {2022})}\BibitemShut {NoStop}%
\bibitem [{\citenamefont {Zhu}\ \emph {et~al.}(2024)\citenamefont {Zhu}, \citenamefont {Chen}, \citenamefont {Liu} \emph {et~al.}}]{Observation07023w}%
  \BibitemOpen
  \bibfield  {author} {\bibinfo {author} {\bibfnamefont {Y.-P.}\ \bibnamefont {Zhu}}, \bibinfo {author} {\bibfnamefont {X.}~\bibnamefont {Chen}}, \bibinfo {author} {\bibfnamefont {X.-R.}\ \bibnamefont {Liu}}, \emph {et~al.},\ }\bibfield  {title} {\bibinfo {title} {Observation of plaid-like spin splitting in a noncoplanar antiferromagnet},\ }\href {https://doi.org/10.1038/s41586-024-07023-w} {\bibfield  {journal} {\bibinfo  {journal} {Nature}\ }\textbf {\bibinfo {volume} {626}},\ \bibinfo {pages} {523} (\bibinfo {year} {2024})}\BibitemShut {NoStop}%
\bibitem [{\citenamefont {Krempasky}\ \emph {et~al.}(2024)\citenamefont {Krempasky}, \citenamefont {Smejkal}, \citenamefont {D'Souza} \emph {et~al.}}]{Alter10.1038}%
  \BibitemOpen
  \bibfield  {author} {\bibinfo {author} {\bibfnamefont {J.}~\bibnamefont {Krempasky}}, \bibinfo {author} {\bibfnamefont {L.}~\bibnamefont {Smejkal}}, \bibinfo {author} {\bibfnamefont {S.~W.}\ \bibnamefont {D'Souza}}, \emph {et~al.},\ }\bibfield  {title} {\bibinfo {title} {Altermagnetic lifting of {Kramers} spin degeneracy},\ }\href {https://doi.org/10.1038/s41586-023-06907-7} {\bibfield  {journal} {\bibinfo  {journal} {Nature}\ }\textbf {\bibinfo {volume} {626}},\ \bibinfo {pages} {517} (\bibinfo {year} {2024})}\BibitemShut {NoStop}%
\bibitem [{\citenamefont {Feng}\ \emph {et~al.}(2022)\citenamefont {Feng}, \citenamefont {Zhou}, \citenamefont {Smejkal} \emph {et~al.}}]{AHC00866z}%
  \BibitemOpen
  \bibfield  {author} {\bibinfo {author} {\bibfnamefont {Z.}~\bibnamefont {Feng}}, \bibinfo {author} {\bibfnamefont {X.}~\bibnamefont {Zhou}}, \bibinfo {author} {\bibnamefont {Smejkal}}, \emph {et~al.},\ }\bibfield  {title} {\bibinfo {title} {An anomalous {Hall} effect in altermagnetic ruthenium dioxide},\ }\href {https://doi.org/10.1038/s41928-022-00866-z} {\bibfield  {journal} {\bibinfo  {journal} {Nat. Electron}\ }\textbf {\bibinfo {volume} {5}},\ \bibinfo {pages} {735} (\bibinfo {year} {2022})}\BibitemShut {NoStop}%
\bibitem [{\citenamefont {Šmejkal}\ \emph {et~al.}(2020)\citenamefont {Šmejkal}, \citenamefont {González-Hernández}, \citenamefont {Jungwirth},\ and\ \citenamefont {Sinova}}]{sciadv.aaz8809}%
  \BibitemOpen
  \bibfield  {author} {\bibinfo {author} {\bibfnamefont {L.}~\bibnamefont {Šmejkal}}, \bibinfo {author} {\bibfnamefont {R.}~\bibnamefont {González-Hernández}}, \bibinfo {author} {\bibfnamefont {T.}~\bibnamefont {Jungwirth}},\ and\ \bibinfo {author} {\bibfnamefont {J.}~\bibnamefont {Sinova}},\ }\bibfield  {title} {\bibinfo {title} {Crystal time-reversal symmetry breaking and spontaneous {Hall} effect in collinear antiferromagnets},\ }\href {https://doi.org/10.1126/sciadv.aaz8809} {\bibfield  {journal} {\bibinfo  {journal} {Sci. Adv.}\ }\textbf {\bibinfo {volume} {6}},\ \bibinfo {pages} {eaaz8809} (\bibinfo {year} {2020})}\BibitemShut {NoStop}%
\bibitem [{\citenamefont {Antonenko}\ \emph {et~al.}(2025)\citenamefont {Antonenko}, \citenamefont {Fernandes},\ and\ \citenamefont {Venderbos}}]{PhysRevLett.134.096703}%
  \BibitemOpen
  \bibfield  {author} {\bibinfo {author} {\bibfnamefont {D.~S.}\ \bibnamefont {Antonenko}}, \bibinfo {author} {\bibfnamefont {R.~M.}\ \bibnamefont {Fernandes}},\ and\ \bibinfo {author} {\bibfnamefont {J.~W.~F.}\ \bibnamefont {Venderbos}},\ }\bibfield  {title} {\bibinfo {title} {Mirror chern bands and weyl nodal loops in altermagnets},\ }\href {https://doi.org/10.1103/PhysRevLett.134.096703} {\bibfield  {journal} {\bibinfo  {journal} {Phys. Rev. Lett.}\ }\textbf {\bibinfo {volume} {134}},\ \bibinfo {pages} {096703} (\bibinfo {year} {2025})}\BibitemShut {NoStop}%
\bibitem [{\citenamefont {Duan}\ \emph {et~al.}(2025)\citenamefont {Duan}, \citenamefont {Zhang}, \citenamefont {Zhu}, \citenamefont {Liu}, \citenamefont {Zhang}, \citenamefont {\ifmmode \check{Z}\else \v{Z}\fi{}uti\ifmmode~\acute{c}\else \'{c}\fi{}},\ and\ \citenamefont {Zhou}}]{PhysRevLett.134.106801}%
  \BibitemOpen
  \bibfield  {author} {\bibinfo {author} {\bibfnamefont {X.}~\bibnamefont {Duan}}, \bibinfo {author} {\bibfnamefont {J.}~\bibnamefont {Zhang}}, \bibinfo {author} {\bibfnamefont {Z.}~\bibnamefont {Zhu}}, \bibinfo {author} {\bibfnamefont {Y.}~\bibnamefont {Liu}}, \bibinfo {author} {\bibfnamefont {Z.}~\bibnamefont {Zhang}}, \bibinfo {author} {\bibfnamefont {I.}~\bibnamefont {\ifmmode \check{Z}\else \v{Z}\fi{}uti\ifmmode~\acute{c}\else \'{c}\fi{}}},\ and\ \bibinfo {author} {\bibfnamefont {T.}~\bibnamefont {Zhou}},\ }\bibfield  {title} {\bibinfo {title} {Antiferroelectric altermagnets: Antiferroelectricity alters magnets},\ }\href {https://doi.org/10.1103/PhysRevLett.134.106801} {\bibfield  {journal} {\bibinfo  {journal} {Phys. Rev. Lett.}\ }\textbf {\bibinfo {volume} {134}},\ \bibinfo {pages} {106801} (\bibinfo {year} {2025})}\BibitemShut {NoStop}%
\bibitem [{\citenamefont {Lee}\ \emph {et~al.}(2024)\citenamefont {Lee}, \citenamefont {Lee}, \citenamefont {Jung}, \citenamefont {Jung}, \citenamefont {Kim}, \citenamefont {Lee}, \citenamefont {Seok}, \citenamefont {Kim}, \citenamefont {Park}, \citenamefont {\ifmmode~\check{S}\else \v{S}\fi{}mejkal}, \citenamefont {Kang},\ and\ \citenamefont {Kim}}]{PhysRevLett.132.036702}%
  \BibitemOpen
  \bibfield  {author} {\bibinfo {author} {\bibfnamefont {S.}~\bibnamefont {Lee}}, \bibinfo {author} {\bibfnamefont {S.}~\bibnamefont {Lee}}, \bibinfo {author} {\bibfnamefont {S.}~\bibnamefont {Jung}}, \bibinfo {author} {\bibfnamefont {J.}~\bibnamefont {Jung}}, \bibinfo {author} {\bibfnamefont {D.}~\bibnamefont {Kim}}, \bibinfo {author} {\bibfnamefont {Y.}~\bibnamefont {Lee}}, \bibinfo {author} {\bibfnamefont {B.}~\bibnamefont {Seok}}, \bibinfo {author} {\bibfnamefont {J.}~\bibnamefont {Kim}}, \bibinfo {author} {\bibfnamefont {B.~G.}\ \bibnamefont {Park}}, \bibinfo {author} {\bibfnamefont {L.}~\bibnamefont {\ifmmode~\check{S}\else \v{S}\fi{}mejkal}}, \bibinfo {author} {\bibfnamefont {C.-J.}\ \bibnamefont {Kang}},\ and\ \bibinfo {author} {\bibfnamefont {C.}~\bibnamefont {Kim}},\ }\bibfield  {title} {\bibinfo {title} {Broken kramers degeneracy in altermagnetic {MnTe}},\ }\href {https://doi.org/10.1103/PhysRevLett.132.036702} {\bibfield  {journal} {\bibinfo  {journal} {Phys. Rev. Lett.}\ }\textbf {\bibinfo
  {volume} {132}},\ \bibinfo {pages} {036702} (\bibinfo {year} {2024})}\BibitemShut {NoStop}%
\bibitem [{\citenamefont {Zhou}\ \emph {et~al.}(2024)\citenamefont {Zhou}, \citenamefont {Feng}, \citenamefont {Zhang}, \citenamefont {\ifmmode~\check{S}\else \v{S}\fi{}mejkal}, \citenamefont {Sinova}, \citenamefont {Mokrousov},\ and\ \citenamefont {Yao}}]{PhysRevLett.132.056701}%
  \BibitemOpen
  \bibfield  {author} {\bibinfo {author} {\bibfnamefont {X.}~\bibnamefont {Zhou}}, \bibinfo {author} {\bibfnamefont {W.}~\bibnamefont {Feng}}, \bibinfo {author} {\bibfnamefont {R.-W.}\ \bibnamefont {Zhang}}, \bibinfo {author} {\bibfnamefont {L.}~\bibnamefont {\ifmmode~\check{S}\else \v{S}\fi{}mejkal}}, \bibinfo {author} {\bibfnamefont {J.}~\bibnamefont {Sinova}}, \bibinfo {author} {\bibfnamefont {Y.}~\bibnamefont {Mokrousov}},\ and\ \bibinfo {author} {\bibfnamefont {Y.}~\bibnamefont {Yao}},\ }\bibfield  {title} {\bibinfo {title} {Crystal thermal transport in altermagnetic ${\mathrm{ruo}}_{2}$},\ }\href {https://doi.org/10.1103/PhysRevLett.132.056701} {\bibfield  {journal} {\bibinfo  {journal} {Phys. Rev. Lett.}\ }\textbf {\bibinfo {volume} {132}},\ \bibinfo {pages} {056701} (\bibinfo {year} {2024})}\BibitemShut {NoStop}%
\bibitem [{\citenamefont {Bai}\ \emph {et~al.}(2024)\citenamefont {Bai}, \citenamefont {Feng}, \citenamefont {Liu}, \citenamefont {Šmejkal}, \citenamefont {Mokrousov},\ and\ \citenamefont {Yao}}]{adfm.202409327}%
  \BibitemOpen
  \bibfield  {author} {\bibinfo {author} {\bibfnamefont {L.}~\bibnamefont {Bai}}, \bibinfo {author} {\bibfnamefont {W.}~\bibnamefont {Feng}}, \bibinfo {author} {\bibfnamefont {S.}~\bibnamefont {Liu}}, \bibinfo {author} {\bibfnamefont {L.}~\bibnamefont {Šmejkal}}, \bibinfo {author} {\bibfnamefont {Y.}~\bibnamefont {Mokrousov}},\ and\ \bibinfo {author} {\bibfnamefont {Y.}~\bibnamefont {Yao}},\ }\bibfield  {title} {\bibinfo {title} {Altermagnetism: Exploring new frontiers in magnetism and spintronics},\ }\href {https://doi.org/10.1002/adfm.202409327} {\bibfield  {journal} {\bibinfo  {journal} {Adv. Func. Mater.}\ }\textbf {\bibinfo {volume} {34}},\ \bibinfo {pages} {2409327} (\bibinfo {year} {2024})}\BibitemShut {NoStop}%
\bibitem [{\citenamefont {Song}\ \emph {et~al.}(2025)\citenamefont {Song}, \citenamefont {Bai}, \citenamefont {Zhou}, \citenamefont {Han}, \citenamefont {Reichlova}, \citenamefont {Dil}, \citenamefont {Liu}, \citenamefont {Chen},\ and\ \citenamefont {Pan}}]{NRM10.1038}%
  \BibitemOpen
  \bibfield  {author} {\bibinfo {author} {\bibfnamefont {C.}~\bibnamefont {Song}}, \bibinfo {author} {\bibfnamefont {H.}~\bibnamefont {Bai}}, \bibinfo {author} {\bibfnamefont {Z.}~\bibnamefont {Zhou}}, \bibinfo {author} {\bibfnamefont {L.}~\bibnamefont {Han}}, \bibinfo {author} {\bibfnamefont {H.}~\bibnamefont {Reichlova}}, \bibinfo {author} {\bibfnamefont {J.~H.}\ \bibnamefont {Dil}}, \bibinfo {author} {\bibfnamefont {J.}~\bibnamefont {Liu}}, \bibinfo {author} {\bibfnamefont {X.}~\bibnamefont {Chen}},\ and\ \bibinfo {author} {\bibfnamefont {F.}~\bibnamefont {Pan}},\ }\bibfield  {title} {\bibinfo {title} {Altermagnets as a new class of functional materials},\ }\href {https://doi.org/10.1038/s41578-025-00779-1} {\bibfield  {journal} {\bibinfo  {journal} {Nat. Rev. Mater.}\ }\textbf {\bibinfo {volume} {10}},\ \bibinfo {pages} {473} (\bibinfo {year} {2025})}\BibitemShut {NoStop}%
\bibitem [{\citenamefont {Zhang}\ \emph {et~al.}(2025{\natexlab{a}})\citenamefont {Zhang}, \citenamefont {Cui}, \citenamefont {Wang}, \citenamefont {Duan}, \citenamefont {Yu},\ and\ \citenamefont {Yao}}]{zhang2025}%
  \BibitemOpen
  \bibfield  {author} {\bibinfo {author} {\bibfnamefont {R.-W.}\ \bibnamefont {Zhang}}, \bibinfo {author} {\bibfnamefont {C.}~\bibnamefont {Cui}}, \bibinfo {author} {\bibfnamefont {Y.}~\bibnamefont {Wang}}, \bibinfo {author} {\bibfnamefont {J.}~\bibnamefont {Duan}}, \bibinfo {author} {\bibfnamefont {Z.-M.}\ \bibnamefont {Yu}},\ and\ \bibinfo {author} {\bibfnamefont {Y.}~\bibnamefont {Yao}},\ }\href@noop {} {\bibinfo {title} {Quantized spin-hall conductivity in altermagnet {Fe$_2$Te$_2$O} with mirror-spin coupling}} (\bibinfo {year} {2025}{\natexlab{a}}),\ \Eprint {https://arxiv.org/abs/2503.10681} {arXiv:2503.10681} \BibitemShut {NoStop}%
\bibitem [{\citenamefont {Feng}\ \emph {et~al.}(2025)\citenamefont {Feng}, \citenamefont {Tan}, \citenamefont {Gao}, \citenamefont {Yan}, \citenamefont {Liu}, \citenamefont {Guo}, \citenamefont {Ma},\ and\ \citenamefont {Lu}}]{feng2025}%
  \BibitemOpen
  \bibfield  {author} {\bibinfo {author} {\bibfnamefont {P.}~\bibnamefont {Feng}}, \bibinfo {author} {\bibfnamefont {C.-Y.}\ \bibnamefont {Tan}}, \bibinfo {author} {\bibfnamefont {M.}~\bibnamefont {Gao}}, \bibinfo {author} {\bibfnamefont {X.-W.}\ \bibnamefont {Yan}}, \bibinfo {author} {\bibfnamefont {Z.-X.}\ \bibnamefont {Liu}}, \bibinfo {author} {\bibfnamefont {P.-J.}\ \bibnamefont {Guo}}, \bibinfo {author} {\bibfnamefont {F.}~\bibnamefont {Ma}},\ and\ \bibinfo {author} {\bibfnamefont {Z.-Y.}\ \bibnamefont {Lu}},\ }\href@noop {} {\bibinfo {title} {{Type-II} quantum spin hall insulator}} (\bibinfo {year} {2025}),\ \Eprint {https://arxiv.org/abs/2503.13397} {arXiv:2503.13397} \BibitemShut {NoStop}%
\bibitem [{\citenamefont {Gonz\'alez-Hern\'andez}\ \emph {et~al.}(2025)\citenamefont {Gonz\'alez-Hern\'andez}, \citenamefont {Serrano},\ and\ \citenamefont {Uribe}}]{PhysRevB.111.085127}%
  \BibitemOpen
  \bibfield  {author} {\bibinfo {author} {\bibfnamefont {R.}~\bibnamefont {Gonz\'alez-Hern\'andez}}, \bibinfo {author} {\bibfnamefont {H.}~\bibnamefont {Serrano}},\ and\ \bibinfo {author} {\bibfnamefont {B.}~\bibnamefont {Uribe}},\ }\bibfield  {title} {\bibinfo {title} {Spin {Chern} number in altermagnets},\ }\href {https://doi.org/10.1103/PhysRevB.111.085127} {\bibfield  {journal} {\bibinfo  {journal} {Phys. Rev. B}\ }\textbf {\bibinfo {volume} {111}},\ \bibinfo {pages} {085127} (\bibinfo {year} {2025})}\BibitemShut {NoStop}%
\bibitem [{SM()}]{SM}%
  \BibitemOpen
  \href@noop {} {\bibinfo  {journal} {{See Supplemental Material for details of the effective lattice model Hamiltonian for altermagnetic multilayers, computational methods of first-principles calculations, and the results of the topological states in Fe$_2$Se$_2$O bilayer and trilayer films}}\ }\BibitemShut {NoStop}%
\bibitem [{\citenamefont {Zhang}\ \emph {et~al.}(2025{\natexlab{b}})\citenamefont {Zhang}, \citenamefont {Cheng}, \citenamefont {Yin}, \citenamefont {Liu} \emph {et~al.}}]{Crystal02864}%
  \BibitemOpen
  \bibfield  {author} {\bibinfo {author} {\bibfnamefont {F.}~\bibnamefont {Zhang}}, \bibinfo {author} {\bibfnamefont {X.}~\bibnamefont {Cheng}}, \bibinfo {author} {\bibfnamefont {Z.}~\bibnamefont {Yin}}, \bibinfo {author} {\bibfnamefont {C.}~\bibnamefont {Liu}}, \emph {et~al.},\ }\bibfield  {title} {\bibinfo {title} {Crystal-symmetry-paired spin-valley locking in a layered room-temperature metallic altermagnet candidate},\ }\href {https://doi.org/10.1038/s41567-025-02864-2} {\bibfield  {journal} {\bibinfo  {journal} {Nat. Phys.}\ }\textbf {\bibinfo {volume} {21}} (\bibinfo {year} {2025}{\natexlab{b}})}\BibitemShut {NoStop}%

\end{thebibliography}

\begin{thebibliography}{12}%
\makeatletter
\providecommand \@ifxundefined [1]{%
 \@ifx{#1\undefined}
}%
\providecommand \@ifnum [1]{%
 \ifnum #1\expandafter \@firstoftwo
 \else \expandafter \@secondoftwo
 \fi
}%
\providecommand \@ifx [1]{%
 \ifx #1\expandafter \@firstoftwo
 \else \expandafter \@secondoftwo
 \fi
}%
\providecommand \natexlab [1]{#1}%
\providecommand \enquote  [1]{``#1''}%
\providecommand \bibnamefont  [1]{#1}%
\providecommand \bibfnamefont [1]{#1}%
\providecommand \citenamefont [1]{#1}%
\providecommand \href@noop [0]{\@secondoftwo}%
\providecommand \href [0]{\begingroup \@sanitize@url \@href}%
\providecommand \@href[1]{\@@startlink{#1}\@@href}%
\providecommand \@@href[1]{\endgroup#1\@@endlink}%
\providecommand \@sanitize@url [0]{\catcode `\\12\catcode `\$12\catcode `\&12\catcode `\#12\catcode `\^12\catcode `\_12\catcode `\%12\relax}%
\providecommand \@@startlink[1]{}%
\providecommand \@@endlink[0]{}%
\providecommand \url  [0]{\begingroup\@sanitize@url \@url }%
\providecommand \@url [1]{\endgroup\@href {#1}{\urlprefix }}%
\providecommand \urlprefix  [0]{URL }%
\providecommand \Eprint [0]{\href }%
\providecommand \doibase [0]{https://doi.org/}%
\providecommand \selectlanguage [0]{\@gobble}%
\providecommand \bibinfo  [0]{\@secondoftwo}%
\providecommand \bibfield  [0]{\@secondoftwo}%
\providecommand \translation [1]{[#1]}%
\providecommand \BibitemOpen [0]{}%
\providecommand \bibitemStop [0]{}%
\providecommand \bibitemNoStop [0]{.\EOS\space}%
\providecommand \EOS [0]{\spacefactor3000\relax}%
\providecommand \BibitemShut  [1]{\csname bibitem#1\endcsname}%
\let\auto@bib@innerbib\@empty
%</preamble>
\bibitem [{\citenamefont {Hohenberg}\ and\ \citenamefont {Kohn}(1964)}]{pub.1060429813}%
  \BibitemOpen
  \bibfield  {author} {\bibinfo {author} {\bibfnamefont {P.}~\bibnamefont {Hohenberg}}\ and\ \bibinfo {author} {\bibfnamefont {W.}~\bibnamefont {Kohn}},\ }\href {https://doi.org/10.1103/physrev.136.B864} {\bibfield  {journal} {\bibinfo  {journal} {Phys. Rev.}\ }\textbf {\bibinfo {volume} {136}},\ \bibinfo {pages} {B864} (\bibinfo {year} {1964})}\BibitemShut {NoStop}%
\bibitem [{\citenamefont {Kohn}\ and\ \citenamefont {Sham}(1965)}]{pub.1060431417}%
  \BibitemOpen
  \bibfield  {author} {\bibinfo {author} {\bibfnamefont {W.}~\bibnamefont {Kohn}}\ and\ \bibinfo {author} {\bibfnamefont {L.~J.}\ \bibnamefont {Sham}},\ }\href {https://doi.org/10.1103/physrev.140.A1133} {\bibfield  {journal} {\bibinfo  {journal} {Phys. Rev.}\ }\textbf {\bibinfo {volume} {140}},\ \bibinfo {pages} {A1133} (\bibinfo {year} {1965})}\BibitemShut {NoStop}%
\bibitem [{\citenamefont {Kresse}\ and\ \citenamefont {Furthmüller}(1996{\natexlab{a}})}]{pub.1060581262}%
  \BibitemOpen
  \bibfield  {author} {\bibinfo {author} {\bibfnamefont {G.}~\bibnamefont {Kresse}}\ and\ \bibinfo {author} {\bibfnamefont {J.}~\bibnamefont {Furthmüller}},\ }\href {https://doi.org/10.1103/physrevb.54.11169} {\bibfield  {journal} {\bibinfo  {journal} {Phys. Rev. B}\ }\textbf {\bibinfo {volume} {54}},\ \bibinfo {pages} {11169} (\bibinfo {year} {1996}{\natexlab{a}})}\BibitemShut {NoStop}%
\bibitem [{\citenamefont {Kresse}\ and\ \citenamefont {Furthmüller}(1996{\natexlab{b}})}]{KRESSE199615}%
  \BibitemOpen
  \bibfield  {author} {\bibinfo {author} {\bibfnamefont {G.}~\bibnamefont {Kresse}}\ and\ \bibinfo {author} {\bibfnamefont {J.}~\bibnamefont {Furthmüller}},\ }\href {https://doi.org/10.1016/0927-0256(96)00008-0} {\bibfield  {journal} {\bibinfo  {journal} {Comput. Mater. Sci.}\ }\textbf {\bibinfo {volume} {6}},\ \bibinfo {pages} {15} (\bibinfo {year} {1996}{\natexlab{b}})}\BibitemShut {NoStop}%
\bibitem [{\citenamefont {Perdew}\ \emph {et~al.}(1996)\citenamefont {Perdew}, \citenamefont {Burke},\ and\ \citenamefont {Ernzerhof}}]{pub.1060814179}%
  \BibitemOpen
  \bibfield  {author} {\bibinfo {author} {\bibfnamefont {J.~P.}\ \bibnamefont {Perdew}}, \bibinfo {author} {\bibfnamefont {K.}~\bibnamefont {Burke}},\ and\ \bibinfo {author} {\bibfnamefont {M.}~\bibnamefont {Ernzerhof}},\ }\href {https://doi.org/10.1103/physrevlett.77.3865} {\bibfield  {journal} {\bibinfo  {journal} {Phys. Rev. Lett.}\ }\textbf {\bibinfo {volume} {77}},\ \bibinfo {pages} {3865} (\bibinfo {year} {1996})}\BibitemShut {NoStop}%
\bibitem [{\citenamefont {Kresse}\ and\ \citenamefont {Joubert}(1999)}]{pub.1060591374}%
  \BibitemOpen
  \bibfield  {author} {\bibinfo {author} {\bibfnamefont {G.}~\bibnamefont {Kresse}}\ and\ \bibinfo {author} {\bibfnamefont {D.}~\bibnamefont {Joubert}},\ }\href {https://doi.org/10.1103/physrevb.59.1758} {\bibfield  {journal} {\bibinfo  {journal} {Phys. Rev. B}\ }\textbf {\bibinfo {volume} {59}},\ \bibinfo {pages} {1758} (\bibinfo {year} {1999})}\BibitemShut {NoStop}%
\bibitem [{\citenamefont {Monkhorst}\ and\ \citenamefont {Pack}(1976)}]{pub.1060521190}%
  \BibitemOpen
  \bibfield  {author} {\bibinfo {author} {\bibfnamefont {H.~J.}\ \bibnamefont {Monkhorst}}\ and\ \bibinfo {author} {\bibfnamefont {J.~D.}\ \bibnamefont {Pack}},\ }\href {https://doi.org/10.1103/physrevb.13.5188} {\bibfield  {journal} {\bibinfo  {journal} {Phys. Rev. B}\ }\textbf {\bibinfo {volume} {13}},\ \bibinfo {pages} {5188} (\bibinfo {year} {1976})}\BibitemShut {NoStop}%
\bibitem [{\citenamefont {Grimme}(2006)}]{jcc.20495}%
  \BibitemOpen
  \bibfield  {author} {\bibinfo {author} {\bibfnamefont {S.}~\bibnamefont {Grimme}},\ }\href {https://onlinelibrary.wiley.com/doi/abs/10.1002/jcc.20495} {\bibfield  {journal} {\bibinfo  {journal} {J. Comput. Chem.}\ }\textbf {\bibinfo {volume} {27}},\ \bibinfo {pages} {1787} (\bibinfo {year} {2006})}\BibitemShut {NoStop}%
\bibitem [{\citenamefont {Aryasetiawan}\ \emph {et~al.}(2006)\citenamefont {Aryasetiawan}, \citenamefont {Karlsson}, \citenamefont {Jepsen},\ and\ \citenamefont {Schonberger}}]{pub.1024097612}%
  \BibitemOpen
  \bibfield  {author} {\bibinfo {author} {\bibfnamefont {F.}~\bibnamefont {Aryasetiawan}}, \bibinfo {author} {\bibfnamefont {K.}~\bibnamefont {Karlsson}}, \bibinfo {author} {\bibfnamefont {O.}~\bibnamefont {Jepsen}},\ and\ \bibinfo {author} {\bibfnamefont {U.}~\bibnamefont {Schonberger}},\ }\href {https://doi.org/10.1103/physrevb.74.125106} {\bibfield  {journal} {\bibinfo  {journal} {Phys. Rev. B}\ }\textbf {\bibinfo {volume} {74}},\ \bibinfo {pages} {125106} (\bibinfo {year} {2006})}\BibitemShut {NoStop}%
\bibitem [{\citenamefont {Mostofi}\ \emph {et~al.}(2014)\citenamefont {Mostofi}, \citenamefont {Yates}, \citenamefont {Pizzi}, \citenamefont {Lee}, \citenamefont {Souza}, \citenamefont {Vanderbilt},\ and\ \citenamefont {Marzari}}]{MOSTOFI20142309}%
  \BibitemOpen
  \bibfield  {author} {\bibinfo {author} {\bibfnamefont {A.~A.}\ \bibnamefont {Mostofi}}, \bibinfo {author} {\bibfnamefont {J.~R.}\ \bibnamefont {Yates}}, \bibinfo {author} {\bibfnamefont {G.}~\bibnamefont {Pizzi}}, \bibinfo {author} {\bibfnamefont {Y.-S.}\ \bibnamefont {Lee}}, \bibinfo {author} {\bibfnamefont {I.}~\bibnamefont {Souza}}, \bibinfo {author} {\bibfnamefont {D.}~\bibnamefont {Vanderbilt}},\ and\ \bibinfo {author} {\bibfnamefont {N.}~\bibnamefont {Marzari}},\ }\href {https://www.sciencedirect.com/science/article/pii/S001046551400157X} {\bibfield  {journal} {\bibinfo  {journal} {Comput. Phys. Commun.}\ }\textbf {\bibinfo {volume} {185}},\ \bibinfo {pages} {2309} (\bibinfo {year} {2014})}\BibitemShut {NoStop}%
\bibitem [{\citenamefont {Sancho}\ \emph {et~al.}(1985)\citenamefont {Sancho}, \citenamefont {Sancho}, \citenamefont {JMLSancho},\ and\ \citenamefont {JRubio}}]{MPLopezSancho_1985}%
  \BibitemOpen
  \bibfield  {author} {\bibinfo {author} {\bibfnamefont {M.~L.}\ \bibnamefont {Sancho}}, \bibinfo {author} {\bibfnamefont {J.~L.}\ \bibnamefont {Sancho}}, \bibinfo {author} {\bibnamefont {JMLSancho}},\ and\ \bibinfo {author} {\bibnamefont {JRubio}},\ }\href {https://doi.org/10.1088/0305-4608/15/4/009} {\bibfield  {journal} {\bibinfo  {journal} {J. Phys. F: Met. Phys.}\ }\textbf {\bibinfo {volume} {15}},\ \bibinfo {pages} {851} (\bibinfo {year} {1985})}\BibitemShut {NoStop}%
\bibitem [{\citenamefont {Wu}\ \emph {et~al.}(2018)\citenamefont {Wu}, \citenamefont {Zhang}, \citenamefont {Song}, \citenamefont {Troyer},\ and\ \citenamefont {Soluyanov}}]{WU2018405}%
  \BibitemOpen
  \bibfield  {author} {\bibinfo {author} {\bibfnamefont {Q.}~\bibnamefont {Wu}}, \bibinfo {author} {\bibfnamefont {S.}~\bibnamefont {Zhang}}, \bibinfo {author} {\bibfnamefont {H.-F.}\ \bibnamefont {Song}}, \bibinfo {author} {\bibfnamefont {M.}~\bibnamefont {Troyer}},\ and\ \bibinfo {author} {\bibfnamefont {A.~A.}\ \bibnamefont {Soluyanov}},\ }\href {https://www.sciencedirect.com/science/article/pii/S0010465517303442} {\bibfield  {journal} {\bibinfo  {journal} {Comput. Phys. Commun.}\ }\textbf {\bibinfo {volume} {224}},\ \bibinfo {pages} {405} (\bibinfo {year} {2018})}\BibitemShut {NoStop}%
\end{thebibliography}


\end{widetext}

\end{document}